\documentstyle[aaspp4,emulateapj5,apjfonts]{article}

\gdef\hdfs{HDF-S}

\gdef\h50min{$h_{50}^{-1}$}
\gdef\kms{km\,s$^{-1}$}

\gdef\3727{[O\,{\sc ii}]\,3727\,\AA}
\gdef\o4959{[O\,{\sc iii}]\,4959\,\AA}
\gdef\5007{[O\,{\sc iii}]\,5007\,\AA}

\gdef\oiii{[O\,{\sc iii}]}
\gdef\msclus{MS\,1054--03}
\gdef\cdfs{CDF-S}
\gdef\4ang{4000\,\AA}
\gdef\orgsf{distant red galaxies}
\gdef\orgs{DRGs}
\gdef\org{DRG}
\gdef\ms1671{MS-1671}
\lefthead{van Dokkum et al.}
\righthead{NIR spectroscopy of red galaxies}
\slugcomment{Accepted for publication in the Astrophysical Journal}
\begin{document}

\title{Stellar Populations and Kinematics of Red
Galaxies at $z>2$: Implications for the Formation of Massive Galaxies
\altaffilmark{1,2}}

\author{Pieter~G.~van Dokkum\altaffilmark{3,4},
Marijn Franx\altaffilmark{5},
Natascha M.~F\"orster Schreiber\altaffilmark{5},
Garth D.~Illingworth\altaffilmark{6},
Emanuele Daddi\altaffilmark{7},
Kirsten Kraiberg Knudsen\altaffilmark{5},
Ivo Labb\'e\altaffilmark{5},
Alan Moorwood\altaffilmark{7},
Hans-Walter Rix\altaffilmark{8},
Huub R\"ottgering\altaffilmark{5},
Gregory Rudnick\altaffilmark{9},
Ignacio Trujillo\altaffilmark{8},
Paul van der Werf\altaffilmark{5},
Arjen van der Wel\altaffilmark{5},
Lottie van Starkenburg\altaffilmark{5},
and Stijn Wuyts\altaffilmark{5}
}

\altaffiltext{1}
{Based on observations obtained at the W.\ M.\ Keck Observatory,
which is operated jointly by the California Institute of
Technology and the University of California.}
\altaffiltext{2}
{Based on observations collected at the European Southern Observatory,
Paranal, Chile (ESO LP 164.O-0612)}
\altaffiltext{3}{Department of Astronomy, Yale University,
New Haven, CT 06520-8101}
\altaffiltext{4}{Astronomy Department, Caltech MS\,105-24, Pasadena, CA 91125}
\altaffiltext{5}{Leiden Observatory, P.O. Box 9513, NL-2300 RA, Leiden,
The Netherlands}
\altaffiltext{6}{UCO/Lick Observatory, University of California, Santa
Cruz, CA 95064}
\altaffiltext{7}{European Southern Observatory, Karl-Schwarzschild-Str.~2,
D-85748, Garching, Germany}
\altaffiltext{8}{Max-Planck-Institute f\"ur Astronomie, K\"onigstuhl 17,
Heidelberg, Germany}
\altaffiltext{9}{Max-Planck-Institute f\"ur Astrophysik, Karl-Scharzschild-Str.~1,
Garching, Germany}

\begin{abstract}

We recently identified a substantial population of galaxies at $z>2$
with comparatively
red rest-frame optical colors. These distant red galaxies (DRGs)
are efficiently selected by the simple observed color criterion
$J_s-K_s>2.3$.
In this paper we present near-infrared
spectroscopy with Keck/NIRSPEC of six DRGs with previously measured
redshifts $2.4<z<3.2$, two of which were known to host an active
nucleus. We detect continuum emission and emission lines of all
observed galaxies.  Equivalent widths of H$\alpha$ in the non-active
galaxies are 20\,\AA --30\,\AA, smaller than measured for Lyman break
galaxies (LBGs) and
nearby luminous infrared galaxies, and comparable to  normal
nearby galaxies. The modest equivalent widths imply that the galaxies
either have a decreasing star formation rate, or that they are very dusty.
Fitting both the photometry and the H$\alpha$ lines, we 
find continuum extinction $A_V =1-2$\,mag, ages $1 -
2.5$\,Gyr, star formation rates $200-400\,M_{\odot}$\,yr$^{-1}$, and
stellar masses $1-5 \times 10^{11}$\,$M_{\odot}$ for models with
constant star formation rates. Models with a
declining star formation lead to significantly lower extinction,
star formation rates, and ages, but similar stellar masses.
From [N\,{\sc ii}]\,/\,H$\alpha$ ratios we infer that
the metallicities are high, $1-1.5 \times$  Solar.
For four
galaxies we can determine line widths from the optical emission lines.
The widths are high, ranging from $130$\,\kms\ to 240\,\kms, and by
combining data for LBGs and DRGs we find significant correlations between
linewidth and restframe $U-V$ color, and between linewidth and stellar
mass. The latter correlation has a similar slope and offset as the
``baryonic Tully-Fisher relation'' for nearby galaxies.  From the
linewidths and effective radii we infer dynamical masses and
mass-to-light ($M/L$) ratios. The median dynamical mass is $\sim 2
\times 10^{11}\,M_{\odot}$, supporting the high stellar
masses inferred from the photometry.
We find that the median $M/L_V \approx 0.8\,(M/L)_{\odot}$, a
factor of $\sim 5$ higher than measured for LBGs.  We infer from our
small sample that DRGs are dustier, more metal rich, more massive, and
have higher ages than $z\approx 3$ LBGs of the same rest-frame $V$-band luminosity.
Although their volume density is still uncertain, their high $M/L$
ratios imply that they contribute significantly  to the stellar mass
density at $z\approx 2.5$.  As their stellar masses are
comparable to those of early-type galaxies they
may have already assembled most of their final mass.

\end{abstract}


\section{Introduction}

It is still not known how and when massive galaxies were formed.
Early studies envisioned an epoch of rapid collapse at high redshift,
followed by sustained star formation in a disk and a smooth and
regular dimming of the stellar light in the spheroidal component
(e.g., {Eggen}, {Lynden-Bell}, \&  {Sandage} 1962). More recently, hierarchical galaxy
formation models in Cold Dark Matter (CDM) cosmologies have postulated
that massive galaxies 
have complex assembly
histories, and were built up gradually through mergers
and periods of star formation ({White} \& {Frenk} 1991). In these models
properties such as mass, star formation rate, and morphology are
transient, depending largely on the merger history and
the time elapsed since the most recent
merger (e.g., {Kauffmann}, {White}, \&  {Guiderdoni} 1993; {Kauffmann} {et~al.} 1999; {Baugh} {et~al.} 1998; {Meza} {et~al.} 2003).

One of the most direct tests of hierarchical galaxy
formation models is the predicted decline of
the abundance of massive galaxies with redshift. The best studied
galaxies at high redshift are $z\approx 3$ Lyman break galaxies (LBGs)
({Steidel} \& {Hamilton} 1992; {Steidel}, {Pettini}, \&  {Hamilton} 1995; {Steidel} {et~al.} 1996).  Measurements have been
made of their clustering properties ({Giavalisco} {et~al.} 1998), star
formation histories ({Papovich}, {Dickinson}, \&  {Ferguson} 2001; {Shapley} {et~al.} 2001), contribution to the cosmic
star formation rate ({Madau} {et~al.} 1996; {Steidel} {et~al.} 1999; {Adelberger} \& {Steidel} 2000),
rest-frame optical emission lines ({Pettini} {et~al.} 2001), and
interaction with the IGM ({Adelberger} {et~al.} 2003). Initially LBGs were
thought to be very massive ({Steidel} {et~al.} 1996), but these estimates
have been revised following the realization that their rest-frame UV
kinematics are dominated by winds rather than gravitational motions
({Franx} {et~al.} 1997; {Pettini} {et~al.} 1998). The masses of luminous
$z\approx 3$ LBGs appear to
be $\sim 10^{10} M_{\odot}$ ({Pettini} {et~al.} 2001; {Shapley} {et~al.} 2001), a factor of $\sim 10$
lower than the most massive galaxies today. Such relatively
low masses are qualitatively consistent with
hierarchical models. LBGs in these models are
``seeds'' marking the
highest density peaks in the early Universe, and form the low
mass building blocks of massive galaxies in groups and clusters
(e.g., {Baugh} {et~al.} 1998).

Despite these advances and the qualitative consistency of
theory and observations many uncertainties remain. Perhaps most
fundamentally, it is still not clear whether LBGs are the (only)
progenitors of today's massive galaxies.  The highly successful Lyman
break technique selects objects with strong UV emission, corresponding
to galaxies with high star formation rates and a limited amount of
obscuration of the stellar continuum.  Galaxies whose light is
dominated by evolved stellar populations, and those that are heavily
obscured by dust, may therefore be underrepresented in LBG samples
(see, e.g., Francis, Woodgate, \& Danks 1997; Stiavelli et al.\ 2001;
Hall et al.\ 2001; {Blain} {et~al.} 2002; {Franx} {et~al.} 2003).

Recent advances in near-infrared (NIR) capabilities on large
telescopes have made it possible to select high redshift galaxies in
the rest-frame optical rather than the rest-frame UV. The rest-frame
optical is much less sensitive to dust extinction and is expected to
be a better tracer of stellar mass.  The FIRES project
({Franx} {et~al.} 2003) is the deepest ground-based NIR survey to date, with
101.5 hrs of VLT time invested in a single pointing on the Hubble Deep
Field South (HDF-S) ({Labb{\' e}} {et~al.} 2003), and a further 77 hrs on a mosaic
of four pointings centered on the foreground cluster \msclus\ (van
Dokkum et al.\ 2003; F\"orster Schreiber et al.\ 2004).

We have selected high redshift galaxies in the FIRES fields by their
observed NIR colors.  The simple criterion $J_s-K_s>2.3$ efficiently
isolates galaxies with prominent Balmer- or 4000\,\AA-breaks at $z>2$
(see {Franx} {et~al.} 2003).  This rest-frame optical break selection is
complementary to the rest-frame UV Lyman break selection.  We find
large numbers of red $z>2$ objects in both fields
({Franx} {et~al.} 2003; {van Dokkum} {et~al.} 2003). Surface densities are $\sim
3$\,arcmin$^{-2}$ to $K_s=22.5$ and $\sim 1$\,arcmin$^{-2}$ to
$K_s=21$, and the space density is 30--50\,\% of that of LBGs.
Their much redder colors suggest they have
higher mass-to-light ($M/L$) ratios, and they may contribute equally
to the stellar mass density.
Most of the galaxies are too faint in the rest-frame UV to be selected
as LBGs.
Although the samples are still too small for robust
measurements, there are indications that the population is highly
clustered ({Daddi} {et~al.} 2003; van Dokkum et al.\ 2003;
{R{\" o}ttgering} {et~al.} 2003).
The available evidence suggests they could be the
most massive galaxies at high redshift, and progenitors of today's
early-type galaxies.

Although these results are intriguing, large uncertainties remain.
The density and clustering measurements are based on very small areas,
comprising less than five percent of the area surveyed for LBGs by Steidel and
collaborators. Furthermore, owing to their faintness in the observer's
optical, spectroscopic redshifts have been secured for only a handful
of objects (van Dokkum et al.\ 2003). Finally,
current estimates of ages, $M/L$ ratios, star formation rates, and
extinction are solely based on modeling of broad band spectral energy
distributions (SEDs), and as is well known this type of analyis
suffers from significant degeneracies in the fitted parameters
(see, e.g., {Papovich} {et~al.} 2001; {Shapley} {et~al.} 2001).

Confirmation of the high stellar masses and improved constraints on
the stellar populations require spectroscopy in the rest-frame optical
(the observer's NIR). Emission lines such as \oiii\ $\lambda
4959,5007$ and the Balmer H$\alpha$ and H$\beta$ lines have been
studied extensively at low redshift, allowing direct comparisons to
nearby galaxies. The H$\alpha$ line is particularly valuable, as its
luminosity is proportional to the star formation rate
({Kennicutt} 1998), and its equivalent width is sensitive to
the ratio of current and past star formation activity.
When more lines are available metallicity and reddening
can be constrained as well.  Finally,  the widths of
rest-frame optical lines
better reflect the velocity dispersion of the H\,{\sc ii} regions than
the widths of
rest-UV lines, which are very sensitive to outflows and supernova-driven winds
(see, e.g., {Pettini} {et~al.} 1998).

In this paper, we present NIR spectroscopy of a small sample of
$J_s-K_s$ selected galaxies. Line luminosities,
equivalent widths, and linewidths are determined, and
the derived constraints on the stellar populations
and masses are combined with results from fits to the broad band SEDs.
Results are compared to nearby galaxies, and also to LBGs:
{Pettini} {et~al.} (1998, 2001) and
{Erb} {et~al.} (2003) have studied the rest-frame optical
spectra of LBGs in great detail, providing an excellent benchmark
for such comparisons.

For convenience we use the term \orgsf, or \orgs, for galaxies having
$J_s-K_s>2.3$ and redshifts $z\gtrsim 2$. This term is
more general than ``optical-break galaxies'', as it allows for
the possibility that in some galaxies
the red colors are mainly caused by dust rather than
a strong continuum break. {Im} {et~al.} (2002) use the
designation Hyper Extremely Red Objects, or HEROs, for galaxies with
$J-K\gtrsim 2$. However, the corresponding rest-frame optical limits
are not really ``hyper extreme'', as they would include all but the bluest nearby galaxies:
at $z=2.7$, our $J_s-K_s$ limit corresponds to $U-V\gtrsim 0.1$ in the
rest-frame. We use $\Omega_m=0.3$, $\Omega_{\Lambda}=0.7$,
and $H_0=70$\,\kms\,Mpc$^{-1}$ ({Riess} {et~al.} 1998; {Spergel} {et~al.} 2003).
All magnitudes are on the Vega system; AB conversion constants
for the ISAAC filters are given in {Labb{\' e}} {et~al.} (2003).

\section{Spectroscopy}

\subsection{Sample Selection and Observations}

NIR spectroscopy of high redshift galaxies is more efficient
if the redshifts are already known, as that greatly facilitates
the search for emission and absorption features.
Although ``blind'' NIR spectroscopy of \orgs\ should be
feasible, for this initial investigation
we chose to limit the sample to galaxies of known redshift.
By January 2003 (the time of our NIRSPEC observations)
redshifts had been measured from the rest-frame
UV spectra of seven \orgs\ at
$z>2$. Five are in the field of 
the $z=0.83$ cluster \msclus, and are presented in {van Dokkum} {et~al.} (2003).
Two are in the Chandra Deep Field South (CDF-S); the rest-frame UV
spectra of these galaxies are presented in the Appendix.
Of the seven galaxies, two show rest-frame UV lines characteristic of
Active Galactic Nuclei (AGN).

Six of the seven \orgs\ were observed with NIRSPEC (McLean et al.\
1998) on Keck II on the nights of 2003 January 21 -- 24.  The
remaining galaxy was given the lowest priority because it is at $z=2.705$,
which means that H$\alpha$ and H$\beta$ fall outside the NIR
atmospheric windows.
NIRSPEC was used in the medium-dispersion mode, providing a plate
scale of 2.7\,\AA\,pix$^{-1}$ in the $H$ band and
$4.2$\,\AA\,pix$^{-1}$ in the $K$ band. The wavelength range offered
by the $1024 \times 1024$ InSb detector is thus roughly equal to the
width of a single NIR atmospheric window.  The $0\farcs 76$ slit
provides a resolution of 10\,\AA\ FWHM in $H$ and 14.5\,\AA\ in $K$,
corresponding to $\sigma_{\rm instr} \approx
80$\,\kms\ in each band.

The night of January 24 was lost in its entirety due to high winds
and fog.  The weather was variable January 21 -- 23, with strong winds
toward the end of each night and the night of January 22 partly
cloudy.  The seeing was typically $0\farcs 8 - 1\farcs 0$ in $K$.
As conditions varied from mediocre to
bad during our four-night run the data presented here are not
representative of the full capabilities of NIRSPEC.

The targets
were acquired using blind offsets from nearby stars.
After an on-target exposure of 900\,s the offset was reversed
and the star was moved along the slit before offsetting again
to the object.
This method allows
continuous monitoring of slit alignment, which proved to be
critical as significant misalignments between slit and setup
star were sometimes observed (and corrected).
In most cases spectra were obtained at three dither positions
(at $0''$, $+7''$, and $-7''$ with
respect to center).
After each
series of dithered observations a nearby
bright A0 star was observed with the same setup to enable correction
for atmospheric absorption and the detector response function.

Our procedure is similar
to that employed by Pettini et al.\ (2001) and Erb et al.\ (2003)
for LBGs, with one modification: after each reversed offset
we obtained a spectrum of the alignment star.
Although this extra step further increases
already substantial overheads, it is very useful in the
data reduction as it provides the spatial coordinate of the (typically
very faint) object spectrum in each individual 900\,s exposure.

All six galaxies were observed in
the NIRSPEC-7 ($K$) filter. For two galaxies we also obtained
spectra in the NIRSPEC-4 band (covering their Balmer/4000\,\AA\ break
regions) and for one of these two we obtained a NIRSPEC-5
($H$ band) spectrum as well (covering [O\,{\sc iii}] and H$\beta$).
Overheads were approximately a factor of two; the total science integration
time was 10 hours.
Integration times and observed wavelength ranges are listed in Table 1.

\begin{figure*}[t]
\epsfxsize=15cm
\epsffile[-60 14 446 487]{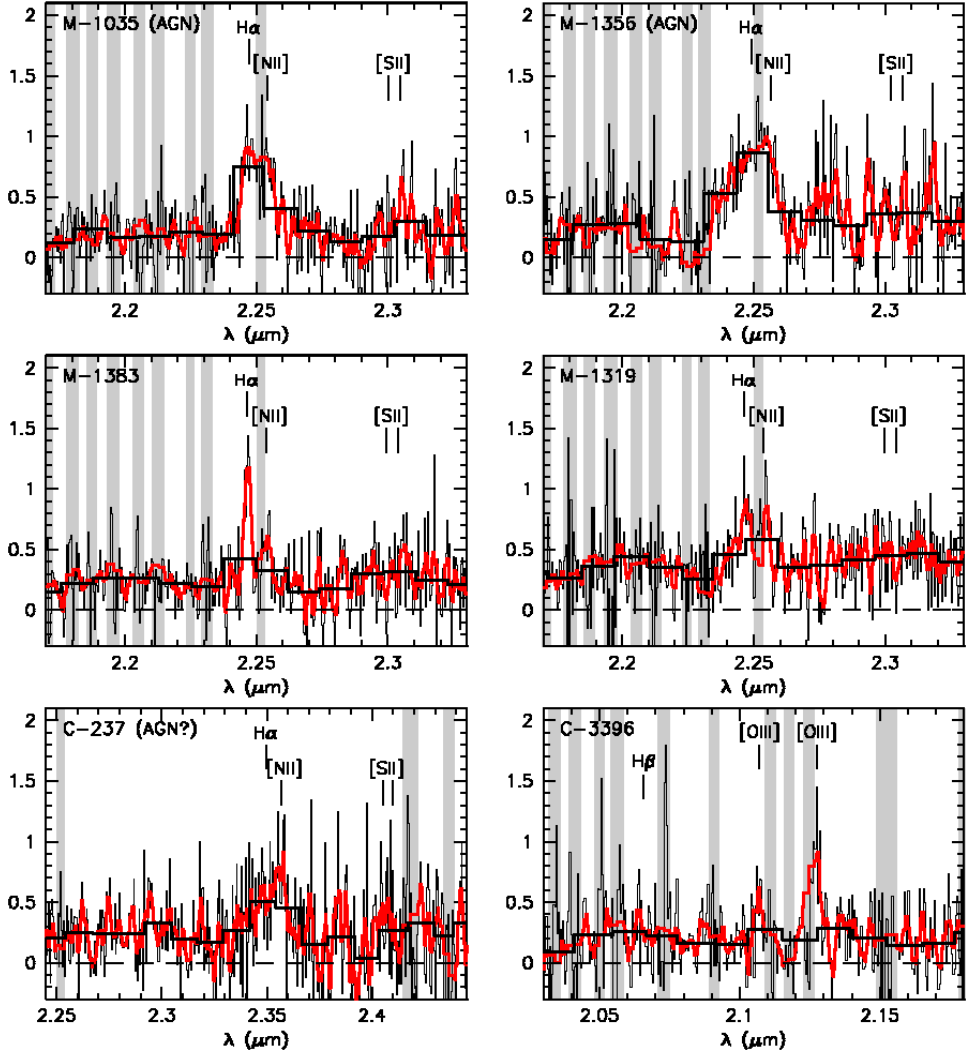}
\caption{\small
NIRSPEC $K$-band spectra of $J_s-K_s$ selected galaxies at
$z>2$. Black
thin lines are raw spectra, black thick histograms are binned spectra, and
smoothed spectra are in red. Grey vertical bands denote areas of
strong sky emission or absorption; these regions were excluded in the
smoothed and binned representations of the data. Emission lines
were detected for all galaxies, with typical rest-frame
equivalent widths of 20--30\,\AA. The continuum is
also detected in every case.
\label{Kspec.plot}}
\end{figure*}

\subsection{Reduction}

The data reduction used a combination of standard IRAF tasks and
custom scripts. As a first step the
strongest cosmic rays were removed, in the following way.
The three individual exposures were averaged while
rejecting the highest value at each
$(x,y)$ position, and the resulting average sky spectrum 
was subtracted from the raw spectra. The
residual spectra are dominated by cosmic rays and sky line residuals resulting
from variations in OH line intensities on $\sim 45$ min timescales.
The residual spectra were rotated by 85.4 degrees
using the IRAF task ``rotate'' such that the sky line residuals were
approximately parallel to columns. Rather than linearly interpolating between
pixels we used ``nearest'' interpolation, so that
each pixel in the rotated frame corresponds exactly to a pixel
in the original frame (with slight redundancy) and cosmic rays retain
their sharp edges. Sky line residuals
were subtracted by fitting a third order polynomial along columns.
Cosmic rays were identified in each of the three residual
images with
{\sc L.A.Cosmic} (van Dokkum 2001), using a 2D noise model created
from the subtracted sky spectrum. The cosmic ray frames were
rotated back to the original orientation
and subtracted from the three raw spectra. 
We carefully checked by eye that sky lines, pixels adjacent to cosmic
rays, and emission lines remained unaffected by the procedure.

Dark current, hot pixels, and a first approximation of the sky
spectrum were removed by subtracting the average of the two other
dither positions from each cosmic ray cleaned
exposure. This procedure increases
the noise in each individual image according to
$
N \approx \sqrt{n/(n-1)} N_{\rm org},
$
with $N_{\rm org}$ the noise before sky subtraction and $n$ the
number of independent dither positions. The number of dither
positions
is a trade-off between this noise increase and
the effects of variations in atmospheric OH line strengths, dark
current, and thermal background on the timescale of the dither
sequence.
For most of our observations $n=3$, and the increase in the
photon noise is $\approx 22$\,\%. 

The frames and their corresponding sky spectra were rotated such that
the sky lines are along columns, using a polynomial to interpolate
between adjacent pixels.  The spectra were wavelength calibrated by
fitting Gaussian profiles to the OH lines in the 2D sky spectra.  The
sky subtracted frames and the sky spectra were rectified to a linear
wavelength scale. Sky line residuals were subtracted by fitting a
second order polynomial along columns, masking the positions of the
object and its two negative counterparts in each frame. As the spectra
of the setup star were reduced in the same way as the science
exposures these positions are known with good accuracy, even though
the object spectra are generally barely visible in individual
exposures.

We extracted two-dimensional 60 pixel ($8\farcs 6$) wide sections
centered on the object from each frame.  Remaining deviant pixels were
removed by flagging pixels deviating more than $4\sigma$ from the
median of the extracted sections.  Finally, the image sections were
averaged, optimally weighting by the signal-to-noise ratio.
Signal-to-noise ratios were determined by averaging the spectra in the
wavelength direction, and fitting Gaussians in the spatial direction.

\subsection{Atmospheric Absorption and Detector Response}

After each 1--2 hr on-target observing sequence we observed
a nearby A0 star with the same instrumental setup. For the \cdfs\
targets SAO\,168314 ($K=9.77$; $J-K=-0.03 \pm 0.03$)
was used, and for targets in
the field of \msclus\ we observed  SAO\,137883 ($K=9.28$;
$J-K=0.01 \pm 0.02$).
The stellar spectra were reduced in the same way as the galaxy
spectra, and divided by the $F_{\lambda}$ spectrum of
Vega. Residuals from Paschen and Brackett absorption lines were
removed by interpolation. The resulting spectra consist of the instrumental
response function and atmospheric absorption.
The galaxy spectra
were divided by these response functions.
No absolute calibration was attempted at this stage, as both
slit losses and the variable
observing conditions would introduce large uncertainties.

\subsection{Extraction of 1D Spectra}

One-dimensional spectra were extracted by averaging all lines
in the 2D spectra with average flux $>0.25 \times$ the average
flux in the central line, using optimal weighting. 
In addition to these averaged ``raw'' spectra, smoothed
and binned spectra were created. Smoothed spectra were
constructed by smoothing the raw 1D spectra with a 5-pixel ($\sim 20$\,\AA)
tophat. Binned spectra sample the raw spectra in 30-pixel ($\sim 120$\,\AA)
bins, with the value in each bin determined with the biweight
estimator ({Beers}, {Flynn}, \& {Gebhardt} 1990). In both cases
pixels at locations of strong sky emission or
strong atmospheric absorption were excluded. 
Extracted $K$-band spectra are shown in Fig.\ \ref{Kspec.plot}.

\section{Photometry}

\subsection{The \msclus\ Field}

The optical and NIR photometry in the \msclus\ field are
described in {van Dokkum} {et~al.} (2003) and
F\"orster Schreiber et al.\ (2004). Briefly, the dataset
is a combination of a mosaic of HST
WFPC2 images ({van Dokkum} {et~al.} 2000) and optical and NIR
data obtained with the VLT as part of the FIRES project
(Franx et al.\ 2000).
The data reach a $5\sigma$ limiting depth of $K_s=23.8$
for point sources. The galaxies described in this paper
have $K_s<20$; therefore, the uncertainties
in the NIR photometry are almost entirely systematic.

The derivation of colors and total $K_s$ magnitudes is
described in F\"orster Schreiber et al.\ (2004). The methodology
follows that of {Labb{\' e}} {et~al.} (2003), who applied identical
procedures to the FIRES dataset on \hdfs.
The observed fluxes required correction for the weak lensing effect of the
foreground $z=0.83$ cluster. The {Hoekstra}, {Franx}, \&  {Kuijken} (2000) weak lensing mass map,
derived from the HST mosaic, was used
to calculate the magnification for each galaxy based on its
position and redshift. The median lensing amplification is 0.15
magnitudes; Table~1 lists the observed and corrected
values. We note that the foreground cluster aided in the
observing efficiency: in its absence we would have had to integrate
20--50\,\% longer to reach the same S/N ratios.  All quantities given
in this paper were derived from the de-lensed magnitudes.

\subsection{Chandra Deep Field South}
\label{cdfphot.sec}

For \cdfs\ we used public VLT/ISAAC NIR observations from the Great Observatories
Origins Deep Survey (GOODS), combined with public European Southern
Observatory (ESO) optical images in the same field. The dataset is
described in {Daddi} {et~al.} (2004). The GOODS imaging
is slightly less deep than the FIRES data in \msclus, but still
more than adequate for the two relatively bright galaxies discussed
here. Magnitudes and colors were measured in a similar way as
in the \msclus\ field. The $H$-band zeropoint is not yet well
determined at the time of writing, as the data have
systematic calibration
problems at the level of 10--15\,\%; we took this into account
by assigning an uncertainty of 0.2 mag to the $H$-band data.
Total $K_s$ magnitudes for both galaxies are listed in Table~1.
Both galaxies have $K_s \leq 20$, but they fall outside the region
studied by the K20 survey ({Cimatti} {et~al.} 2002).

\section{Analysis}

\subsection{Continuum Emission and Spectral Breaks}
\label{balmer.sec}

$K$-band continuum emission is detected of all six
observed galaxies (Fig.\ \ref{Kspec.plot}).  This high detection rate
can be contrasted with results obtained for LBGs by
{Pettini} {et~al.} (2001), who detected continuum emission for only two out
of 16 galaxies in their primary sample. The reason for this difference
is simply that the $K$-magnitudes of our spectroscopic sample of
$J_s-K_s$ objects are typically much brighter than those of
$z \approx 3$ Lyman break galaxies. In fact, their comparitively high luminosity
in the NIR makes the $J_s-K_s$ selected galaxies eminently
suitable for NIR spectroscopy on 8--10m telescopes.
The continuum detections are very important, as they allow us to
determine accurate equivalent widths and line luminosities.
Had the continuum not been detected, absolute calibration of the
spectra would have been required, leading to typical uncertainties of
a factor of two (see {Pettini} {et~al.} 2001).

\vbox{
\begin{center}
\leavevmode
\hbox{%
\epsfxsize=8.5cm
\epsffile{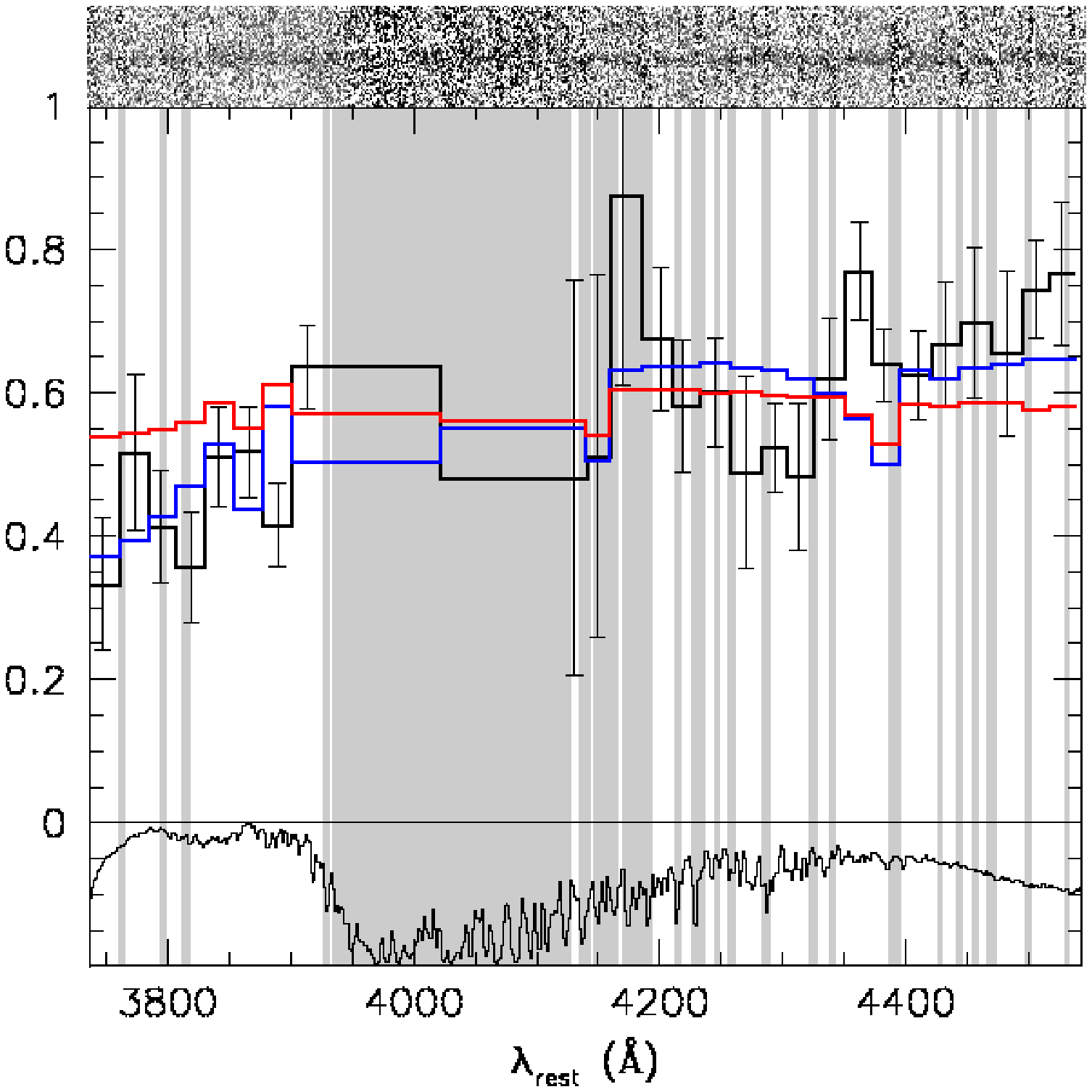}}
\figcaption{\small
Binned spectrum of the brightest galaxy in our sample
(M--1319 at $z=2.43$) in the N4-band (black histogram),
along with the unbinned 2D spectrum (top). Also shown is
the spectrum of a nearby A0 star (bottom) which was used to correct
for atmospheric absorption and detector response. The blue line shows
the Bruzual \& Charlot (2003) model which best fits
the broad band photometry. This model with an age of 2.6\,Gyr and
extinction $A_V=1.3$\,mag provides a
remarkably good fit to the observed spectrum.
The red line shows a model with a very young age and high
extinction ($A_V=3$); this model overpredicts the flux blueward of
3900\,\AA\ (see \S\,\ref{model.sec}).
\label{balmer.plot}}
\end{center}}

The S/N in the continua is not sufficient to detect individual
absorption lines. However, for the brightest galaxy in our
sample we detect a significant drop in the continuum
in the region of the redshifted Balmer/4000\,\AA-break.
Figure \ref{balmer.plot} shows the N-4 spectrum of galaxy
M--1319 ($z=2.423$). The N4-band lies between the $J$ and $H$
bands, and suffers from strong atmospheric absorption.
The continuum on the red side of the absorption trough is
stronger than the continuum on the blue side, indicating a
drop across the Balmer/4000\,\AA-break. The biweight mean
over the wavelength range $1.28-1.33\,\mu$m ($3739-3885$\,\AA\
in the rest-frame) is $0.46 \pm 0.04$ (arbitrary units),
compared to $0.69 \pm 0.03$ over the wavelength range
$1.43-1.57$\,$\mu$m ($4177-4587$\,\AA\
in the rest-frame), a drop of a factor of $\approx 1.5$.

We conclude that the $J_s-K_s$ criterion isolated a galaxy with a
continuum break, as it was designed to do. As will be discussed in
\S\,\ref{model.sec} the S/N is just sufficient
to provide a constraint on the dust content. Moreover,
this detection in a relatively short exposure time
highlights the potential of present-day telescopes
for quantitative continuum studies of
galaxies with evolved stellar populations at $z>2$.

\subsection{Emission Lines}

\subsubsection{Detections}
\label{detect.sec}

Emission lines were detected for all observed galaxies.
Two galaxies were known to host AGNs;
both show strong and fairly broad emission lines. Three of the
remaining galaxies
show narrow lines typical of normal star forming galaxies, and the
fourth shows a broad line probably indicating an active nucleus.
Each of the six observed galaxies is briefly discussed below.

\begin{figure*}[t]
\epsfxsize=15cm
\epsffile[-85 14 432 424]{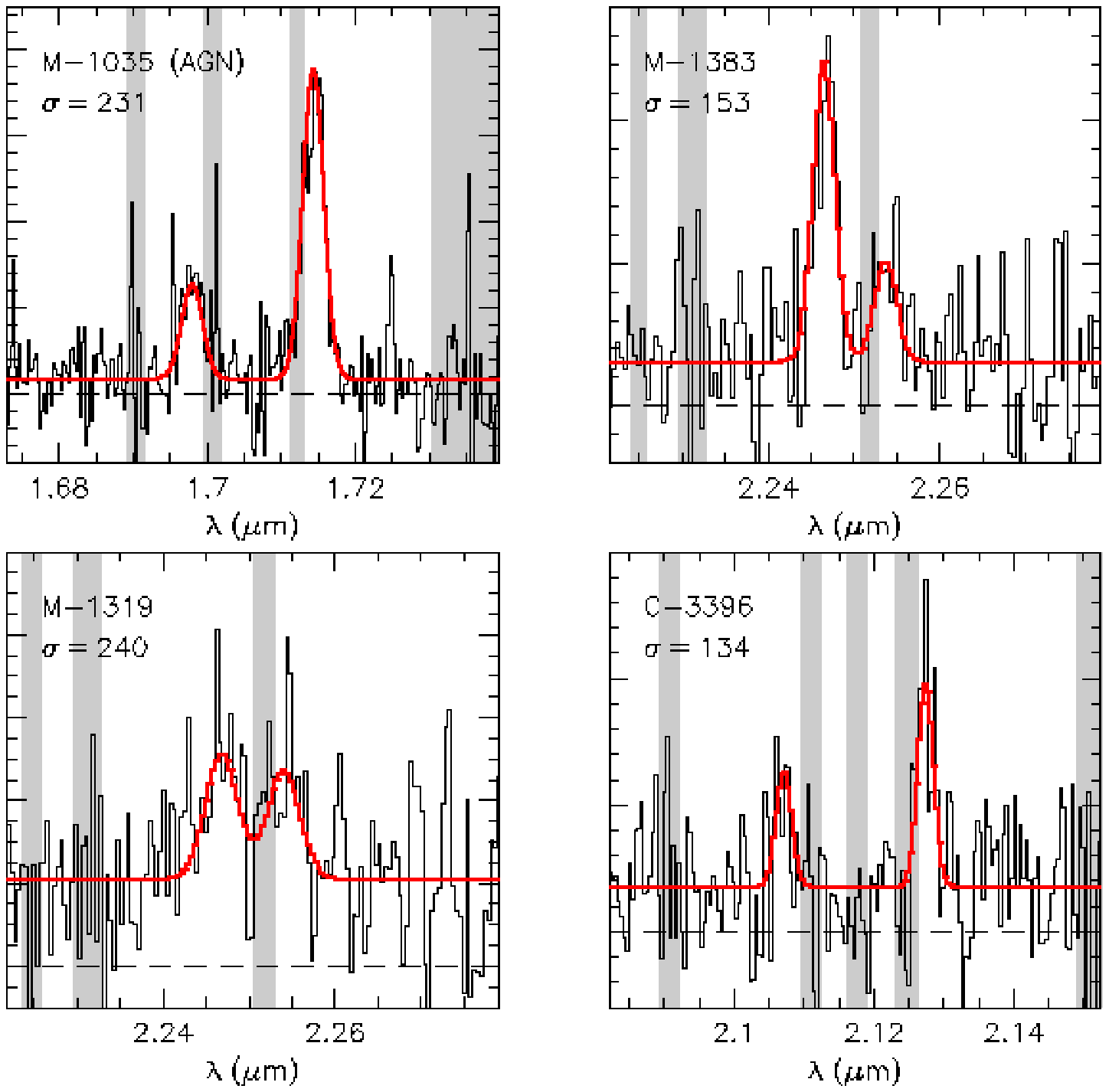}
\caption{\small
Gaussian fits to emission lines. 
For M--1035 and C--3396 the lines
are [O\,{\sc iii}]\,4959,5007\,\AA; for M--1383 and M--1319 they
are H$\alpha$ and [N\,{\sc ii}]\,$\lambda 6584$. Raw spectra
are shown in black; red lines are best fitting models. Line widths
are also shown, corrected for instrumental broadening. The line
widths are substantial, typical of massive nearby galaxies.
\label{kin.plot}}
\end{figure*}

\begin{figure*}[t]
\epsfxsize=16cm
\epsffile[0 185 544 671]{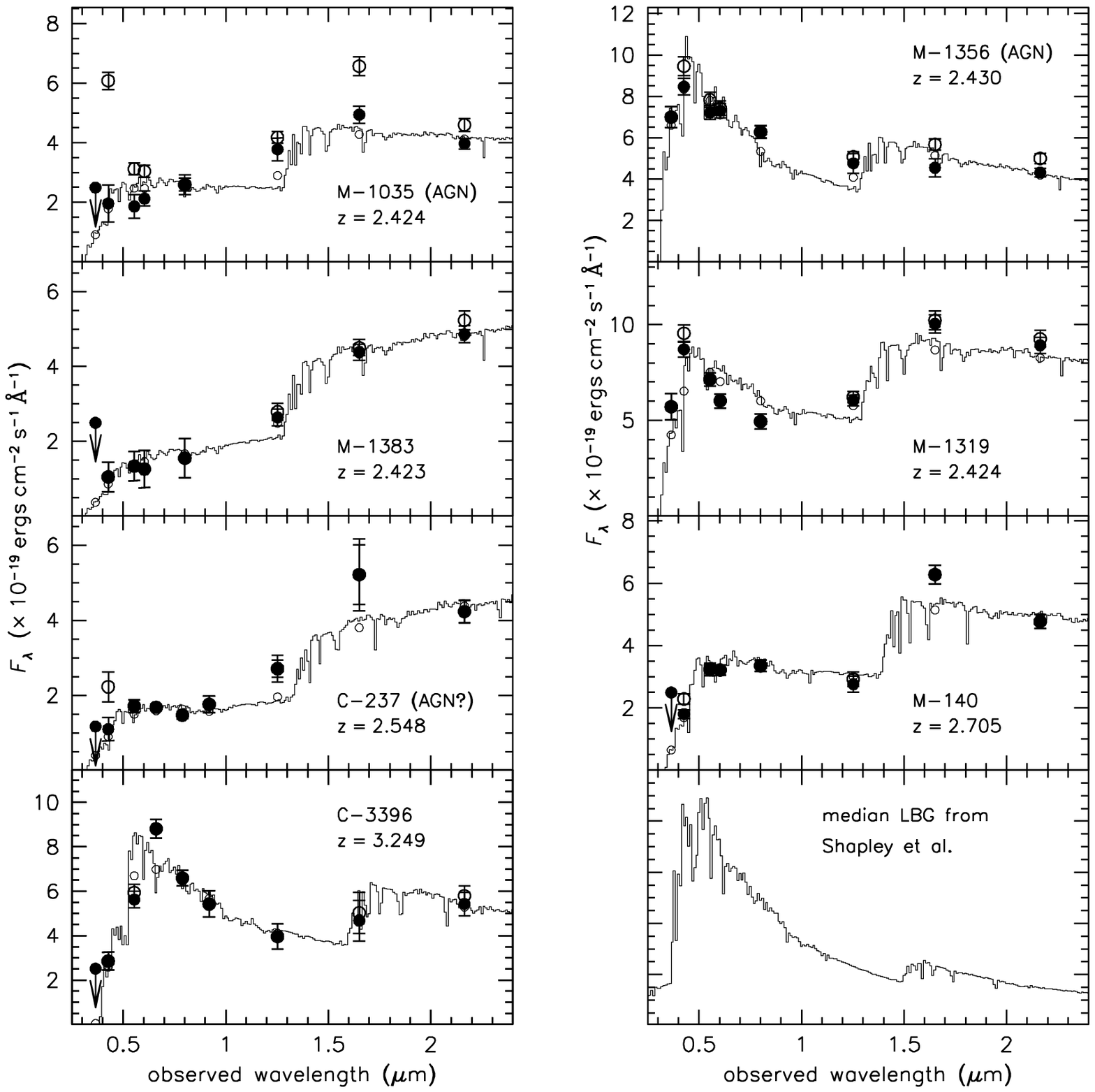}
\caption{\small
Spectral energy distributions of $J_s-K_s$ selected galaxies
with spectroscopic redshifts $z>2$. Large open circles show observed
fluxes, and solid circles show fluxes corrected for
emission line contributions. Thin lines show best fitting
stellar population synthesis models with
constant star formation rate and varying age and dust content.
Small open circles are model fluxes in the broad band filters.
The median age is 1.6\,Gyr and the median absorption
$A_V = 1.8$\,mag.
For comparison the lower right panel shows the median
best fitting model to $z=3$ Lyman break galaxies from
Shapley et al.\ (2001), with a much younger age of 0.3\,Gyr
and absorption $A_V=0.5$\,mag.
\label{seds.plot}}
\end{figure*}

\noindent
{\bf M--1035} --
This object was identified as a Type-II AGN based on its
narrow Ly$\alpha$, C\,{\sc iv} and He\,{\sc ii} lines in the
rest-frame UV ({van Dokkum} {et~al.} 2003). We observed the galaxy in
the N-4, N-5 ($H$), and N-7 ($K$) bands. The continuum was detected
in every band as well as several emission lines.
All lines have widths $<2000$\,\kms,
confirming the Type-II classification (see, e.g., {Stern} {et~al.} 2002; {Norman} {et~al.} 2002). The \oiii\ lines have width $\sigma = 231$\,\kms\
(see \S\,\ref{width.sec}), and
the broadest feature is
the H$\alpha$ + [N\,{\sc ii}] complex, width a {\em combined}
FWHM of 1600\,\kms. The object is a faint hard X-ray source
({Johnson}, {Best}, \& {Almaini} 2003), consistent with the idea that the central
engine is (partially) obscured by dust. 

\noindent
{\bf M--1356} --
The  rest-frame UV spectrum of M--1356 indicates the presence of an AGN
({van Dokkum} {et~al.} 2003),
with characteristics of Broad Absorption Line QSOs.
The UV spectrum shows strong Nitrogen lines indicating
a high metallicity.
The $K$-band spectrum shows fairly broad H$\alpha$ + [N\,{\sc ii}] emission with
FWHM $\sim 2500$\,\kms. This galaxy is not a Chandra source, consistent
with the BAL nature of its active nucleus (e.g., {Green} \& {Mathur} 1996).
We note that due to a revision of our
photometry since the NIRSPEC observations
this galaxy has $J_s-K_s=2.13$, and
hence falls blueward of our \org\ criterion; removing this galaxy
from the analysis does not alter the conclusions in any way.

\noindent
{\bf M--1383} --
For galaxy M--1383 the rest-frame optical spectroscopy implies
a revision of the rest-frame UV redshift.
We deduced a redshift $z=3.525$ from faint absorption lines in the
rest-frame UV ({van Dokkum} {et~al.} 2003); however,
in the $K$-band we detect H$\alpha$ and
[N\,{\sc ii}] redshifted to $z=2.423$, making this the fourth red galaxy
at $z\approx 2.43$ in the \msclus\ field. The revised
redshift is consistent with the photometric redshift in {van Dokkum} {et~al.} (2003).
The rest-UV spectrum has low S/N, and the previously published redshift
may simply be in error. Another possibility is that the blue feature
seen off-center in this object is a
background galaxy at $z=3.52$.

\noindent
{\bf MS--1319} --
The brightest galaxy in the sample in the $K$-band, enabling the
direct detection of the continuum break responsible for its red $J_s-K_s$
color (\S\,\ref{balmer.sec}). The $K$-band spectrum shows
H$\alpha$ and [N\,{\sc ii}].

\noindent
{\bf C--237} -- The spectrum of C--237
shows a broad, low S/N H$\alpha$\,+\,[N\,{\sc ii}] complex.
Higher S/N spectra are needed to measure the width reliably, but
we infer that the galaxy probably hosts an active nucleus.
The association of C--237 with a
hard X-ray source (Alexander et al.\ 2003)
is consistent with this interpretation.

\noindent
{\bf C--3396} --
The redshifted \oiii\ lines are detected at the expected locations;
H$\beta$ is barely detected. The lines are resolved
in the spatial direction, and there is some
evidence for a velocity gradient. The galaxy is a faint soft
source in the 1\,Ms Chandra catalog
(Giacconi et al.\ 2002), possibly indicating a very
strong star burst. The X-ray properties of \orgs\
will be discussed in K.~Rubin et al., in preparation.

\subsubsection{Line Luminosities and Equivalent Widths}
\label{eqwidth.sec}

For each galaxy we modeled the continuum with a second order
polynomial, fitted to the binned spectrum. Emission lines were masked
in the fit. After subtracting the continuum, line fluxes were measured
by summing the counts in the residual spectrum in
regions centered on the redshifted lines. The width of these regions
was typically 50\,\AA, except for two cases where H$\alpha$ is
broader (see \S\,\ref{detect.sec}).  Errors were derived from
simulations, by placing random apertures in regions with similar noise
characteristics and by varying the parameters of the continuum fit.

Calibration of the line fluxes was performed in the following way.
First the emission lines were added to the continuum fits; we used the
fits rather than the measured continuum to limit the effects of sky
line residuals. Next, the spectra were converted from $F_{\lambda}$ to
$F_{\nu}$ and convolved with the VLT ISAAC broad band filter response
curves appropriate for the observed wavelength range.  Finally, the
spectra were normalized using the measured magnitudes in these
filters. The broad band photometry
is accurate to $\approx 0.05$\,mag. Total magnitudes were used, so that the
measured line fluxes are effectively corrected to an aperture covering
the entire galaxy.
Equivalent widths were
determined directly by comparing the line flux to the continuum fit at the
location of the line, and require no absolute calibration.
The measured line properties are listed in Table~2.

\subsubsection{Line Widths}
\label{width.sec}

We determined line widths by fitting Gaussian
models to the 1D spectra. Galaxies C--237 and M--1356 were
excluded, as for both galaxies the
H$\alpha$\,+\,[N{\sc ii}] line complex is too broad to
separate the two lines with confidence.
Galaxy M--1035 also has a broad
H$\alpha$ line complex, but for this object
we can obtain a reliable line width from its \oiii\ lines.

For each of the four galaxies two lines were detected (see
Table~2); both lines were fitted simultaneously.  Free parameters
in the fits are the redshift, fluxes of both
lines, and the line width. The continuum was allowed to vary
within the uncertainties of the polynomical fit to the binned
spectrum (see \S\,\ref{eqwidth.sec}).
The fitting procedure minimizes the
sum of the absolute differences between model and data
to limit the effects of outliers; minimizing
the rms gives similar results. Pixels in regions
of strong sky lines were excluded from the fit.

Best fitting models are shown in Fig.\
\ref{kin.plot}; line widths corrected for instrumental broadening are
listed in Table~2.
Errors were determined from
fits to simulated spectra.
One-dimensional noise spectra were extracted from the 2D
spectra at random spatial positions and added to model spectra; these
artifical spectra have very similar noise characteristics as the
galaxy spectra. The errors were
determined from the biweight width
of the distribution of differences between true and measured line widths.
The line width of galaxy C--3396 has a large (asymmetric)
uncertainty; although the
faint \oiii\ lines are resolved, their
width is not well determined.

\section{Stellar Populations}

\subsection{Fits to Broad Band Photometry}
\label{sedfits.sec}

For all galaxies high quality photometry is available over the
wavelength range $0.35 - 2.2\,\mu$, and following previous studies of LBGs
we fit model SEDs to the
photometric data to constrain star formation rates, ages, and the
extinction of the stellar continua.
For completeness we include galaxy M--140 in this analysis, as
it is the only \org\ with redshift that was not observed with NIRSPEC,
and its redshift implies that the
NIR photometry will only be marginally affected by
emission lines. We note that similar fits
were presented in {van Dokkum} {et~al.} (2003) for the five galaxies in the
\msclus\ field.
Fits to the entire sample of \orgs\ in \msclus\ (i.e., including those
without spectroscopic redshift) are presented in F\"orster Schreiber et al.\ (2004).

\subsubsection{Corrections for Emission Lines}

Strong emission lines can affect the broad
band fluxes of high redshift galaxies and may cause systematic
errors in derived stellar population parameters.
In Lyman break galaxies emission lines typically contribute $\sim 15$\,\% to
the broad band fluxes ({Shapley} {et~al.} 2001).
This estimate is somewhat uncertain
because it relies on absolute calibration of the NIR
spectroscopy (see {Pettini} {et~al.} 2001).
The corrections can be determined with
greater confidence for the present sample of \orgs\
because the continuum is detected in every case.

\begin{figure*}[t]
\epsfxsize=17.5cm
\epsffile{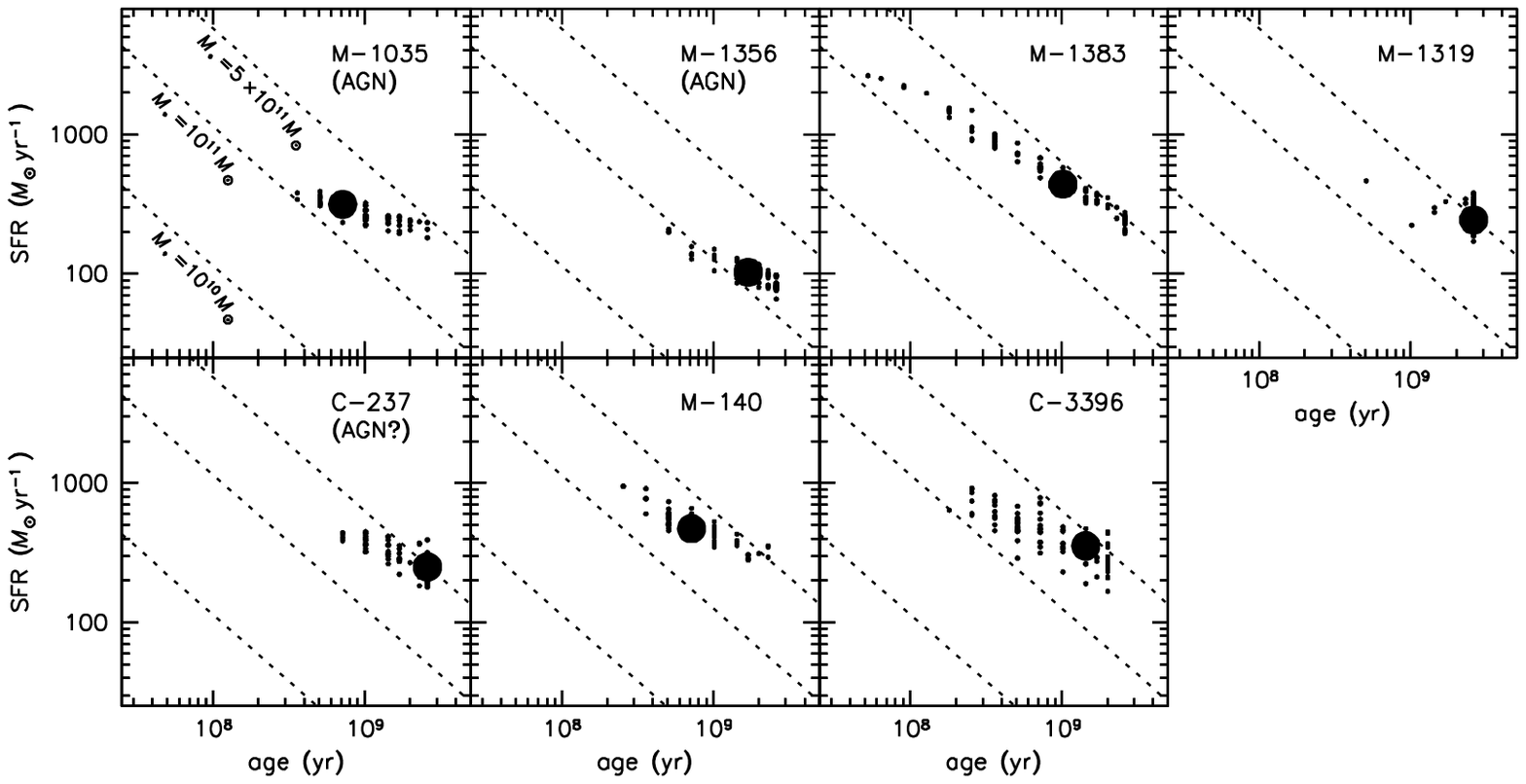}
\caption{\small
Results of Bruzual \& Charlot (2003) model fits to the emission-line corrected
broad-band SEDs. Large symbols show the
best fitting values of age and star formation rate.
Small symbols show the range in best fit parameters for each galaxy,
derived from randomizing the photometry within the (random and
systematic) photometric errors. Dashed lines are of constant mass.
The ages and star formation rates have substantial uncertainties, but
the stellar masses are well constrained. They are very high at
$1-5\,\times\,10^{11}\,M_{\odot}$.
\label{sedresult.plot}}
\end{figure*}

Emission line corrections for all observed photometric bands
are listed in Table~3. Values in parentheses have been
calculated with the help of the following relations,
determined from the
{Jansen} {et~al.} (2000) sample of nearby galaxies:
\begin{equation}
W_{\lambda}({\rm [O}\,{\rm II}{\rm ]}) = 0.98 \pm 0.52
\,\times\,W_{\lambda}({\rm H}\alpha)
\end{equation}
\begin{equation}
W_{\lambda}({\rm H}\beta + {\rm [O}\,{\rm III}{\rm ]}) = 0.53 \pm 0.22
\,\times\,W_{\lambda}({\rm H}\alpha)
\end{equation}
\begin{equation}
W_{\lambda}({\rm [O}\,{\rm II}{\rm ]}) = 1.8 \pm 0.4
\,\times\,W_{\lambda}(\lambda 5007).
\end{equation}
The corrections range from $<0.05$ to $0.16$ magnitudes in the
$K_s$ band, with a median of $0.08$.
Our sample is biased toward the presence of
emission lines in the rest frame UV (as this makes it easier to
measure the redshift), and it is reasonable
to assume that it is also biased toward the presence
of lines in the rest frame optical. Therefore, 
the small contributions measured here
strongly suggest that the red colors of 
$J_s-K_s$ selected galaxies in general
are caused by red
continua rather than the presence of strong emission lines
in the $K_s$ band.


\subsubsection{Fitting Methodology and Results}
\label{model.sec}

Emission-line corrected SEDs are shown in Fig.\ \ref{seds.plot} (solid symbols),
along with the uncorrected photometry (open symbols).
We fitted stellar population
synthesis models to the corrected photometry.
The {Bruzual} \& {Charlot} (2003) models were used;
this latest version is based on a new
library of stellar spectra and has updated prescriptions for
AGB stars.

A major systematic uncertainty in this type of analysis
is the parameterization of the star formation history ({Shapley} {et~al.} 2001; {Papovich} {et~al.} 2001). The ubiquitous emission line detections
in our sample imply that the galaxies are not evolving passively,
and that models in which all stars were formed in a single burst
at high redshift are not appropriate.
For simplicity we limit the discussion
to models with constant star formation rates, characterized by
three parameters: the time since the onset of star formation
(the age), the star formation rate, and the extinction.
Models with a declining star formation rate are discussed briefly
in \S\,\ref{tau.sec}, and in F\"orster Schreiber et al.\ (2004).
We only consider solar metallicity models; as we show in \S\,\ref{Z.sec}
such models are consistent with the limited information we
have on the abundances of \orgs.

The publicly available {\sc hyperz} photometric redshift
code ({Bolzonella}, {Miralles}, \& {Pell{\'  o}} 2000) was used
for the fitting. Its template library was updated with the
{Bruzual} \& {Charlot} (2003) models and the redshift was held fixed
at the spectroscopic value. The {Calzetti} {et~al.} (2000) reddening law
was applied; we tested that the results are not very sensitive to
the assumed extinction curve. A minimum photometric error of
0.05 was assumed, so that individual data points
with very small formal errors do not dominate the
$\chi^2$ minimalization.

Best fitting SEDs are overplotted in Fig.\ \ref{seds.plot}.
Corresponding star formation rates,
ages and reddening values are listed in Table~4.
The constant star formation models provide reasonable descriptions of
the photometric data. The median absolute difference between data and
model is $\approx 0.07$ mag for galaxies C--3396, M--140, M--1356 and
M--1383, and $\approx 0.2$ mag for the other three galaxies.  However, as
the errors in the photometry are typically much smaller, the $\chi^2$
minima are mostly greater than one, with median 2.9.

Confidence limits on the derived parameters were determined in the
following way.  For each galaxy, 500 simulated SED datapoints were created by
randomizing the photometry within the uncertainties. In order to
account for systematic errors and template mismatch the uncertainties
were multiplied by the square root of the $\chi^2$ value of the best
fitting model.  The simulated SEDs were modeled using the same
procedure as the observed SEDs.  Figure\ \ref{sedresult.plot} shows
the best fit age and star formation rate for each galaxy (large
symbols) along with values for 100 of the simulated SEDs (small
symbols).  Uncertainties in the ages are highly correlated with those
in the star formation rates. As stellar mass is the product of star
formation and age in these models the uncertainty in this parameter is
relatively small.

The $1\sigma$ confidence limits listed in Table~4 were determined
from the simulations by calculating the widths of the distributions of
age, star formation rate, reddening, and stellar mass. The confidence
limits are generally not symmetric.  For all galaxies ``maximally
old'' fits fall within the confidence limits, i.e., in every case the
upper limit to the age is set by the age of the universe at the epoch
of observation.  The uncertainties are largest for galaxy M--1383,
which has a smooth SED without a pronounced break.

The stellar populations in these galaxies appear to be quite similar.
All seven have relatively old ages ranging from 0.7 -- 2.6 Gyr, and
substantial extinction ranging from 0.9 -- 2.3 magnitudes in the rest
frame $V$ band. The implied star formation rates range from
102\,$M_{\odot}$\,yr$^{-1}$ to 468\,$M_{\odot}$\,yr$^{-1}$, and
the stellar masses from
$1.3\times 10^{11}\,M_{\odot}$ to $4.9\times
10^{11}\,M_{\odot}$.  We note that the parameters for galaxies
M--1035 and M--1356 are uncertain, as their active nuclei may affect
their continua.

\subsection{Constraints from H$\alpha$}
\label{halpha.sec}

The SED fitting results suggest that the \orgs\ are unlike
Lyman break galaxies: their star formation
rates are similar, but their ages and dust content
are substantially larger. The implied stellar masses
rival those of elliptical galaxies in the nearby Universe.  In light
of these important implications and the systematic uncertainties
associated with SED fitting it is critical to compare the results
to those implied by our NIR spectroscopy.

\subsubsection{Star Formation Rates and Ages}

The H$\alpha$
emission line measures the
current star formation rate (through the line luminosity) as well as
the ratio of current and past star formation (through the equivalent
width). Dust-corrected line
luminosities can be converted to the instantaneous
star formation rate using
\begin{equation}
\label{sfr.eq}
{\rm SFR}\,(M_{\odot}\,{\rm yr}^{-1}) = 7.9 \times 10^{-42} L_{{\rm H}\alpha}
({\rm ergs\,s}^{-1})
\end{equation}
({Kennicutt} 1998). In local galaxies this relation is consistent
with other indicators of the star formation rate, such as radio
luminosity and far-infrared luminosity (e.g., {Kewley} {et~al.} 2002). Such studies have
not yet been done systematically at high redshift, although
results for LBGs suggest that this correspondence holds for these galaxies
as well (e.g., Reddy \& Steidel 2004).

For constant star formation histories the
equivalent width has a one-to-one relation with the time
elapsed since the onset of star formation (the age).
The dependence
of age on equivalent
width has the form of a powerlaw. Using the Bruzual
\& Charlot (2003) models to find the continuum luminosity and
Eq.\ \ref{sfr.eq} to convert line luminosity to star formation rate
we find
\begin{equation}
\label{ew.eq}
\log {\rm (age)} = 13.2 - 2.4\,\log\,W_{\lambda}({\rm H}\alpha)
\end{equation}
with $W_{\lambda}$ the rest frame equivalent width in \AA\ and
the age in years. Over the range $7.5 < \log {\rm (age)} < 9.5$
the powerlaw approximation holds to a few percent. Note that the
relation depends on the conversion factor between
star formation rate and line luminosity (Eq.\ \ref{sfr.eq}),
but not on the star formation rate itself.

The main uncertainty is attenuation by dust. H$\alpha$
luminosities are usually corrected for extinction by comparing
them to H$\beta$, assuming the intrinsic ratio
of case-B recombination. However, as we have not measured
the H$\alpha$/H$\beta$ pair in any of the galaxies, we have to
derive the extinction from the attenuation of the stellar
continuum. Studies of nearby and distant galaxies indicate that
the extinction toward H{\sc ii} regions is at least as high
as that of the stellar continuum (e.g., Kennicutt, 1992, 1998;
Calzetti et al.\ 1996; Erb et al.\ 2003). The relation
depends on the distribution of the dust, which is difficult
to constrain observationally.
In the following, we consider two cases:
1) no additional extinction toward H{\sc ii} regions. In this
case, $L({\rm H}\alpha)_{\rm corr}=10^{0.37\,A_V}L({\rm H}\alpha)_{\rm obs}$
(with $A_V$ the
extinction of the stellar continuum),
and $W_{\lambda}({\rm H}\alpha)_{\rm corr}
= W_{\lambda}({\rm H}\alpha)_{\rm obs}$; and 2) additional $V$-band extinction
of one magnitude. In that case, 
$L({\rm H}\alpha)_{\rm corr}=
10^{0.37\,(A_V+1)} L({\rm H}\alpha)_{\rm obs}$,
and $W_{\lambda}({\rm H}\alpha)_{\rm corr}
= 2.34 W_{\lambda}({\rm H}\alpha)_{\rm obs}$.
These two cases probably bracket the actual values.

\begin{figure*}[t]
\epsfxsize=16.5cm
\epsffile[-50 14 486 417]{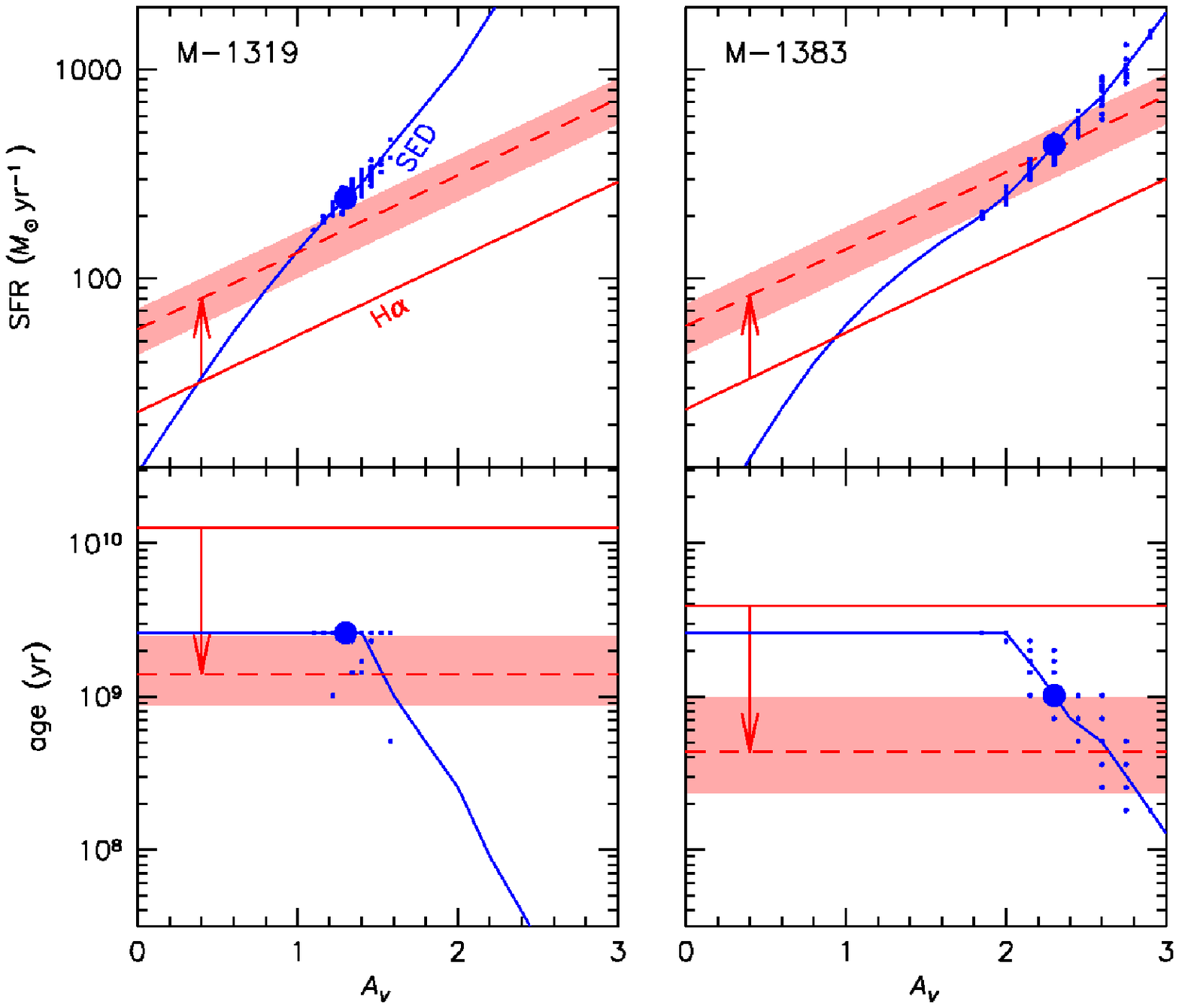}
\caption{\small
Combined constraints on stellar population parameters from SED fitting
(blue) and
H$\alpha$ luminosity and equivalent width (red).
The H$\alpha$ constraints depend on the assumed
extinction toward H\,{\sc ii} regions: the solid line assumes
that it is equal to the extinction of the stellar continuum, and
the dashed line assumes an additional extinction of one magnitude.
The bands indicate 1$\sigma$ observational errors in the H$\alpha$
measurements. Note that the SED constraints on the ages are bounded by the
age of the Universe at the epoch of observation.
Small dots are derived from error simulations
(see Fig.\ \ref{sedresult.plot}).
The H$\alpha$ constraints are less sensitive to
$A_V$ than the SED constraints, and consistent solutions
are obtained where the red and blue lines intersect.
The regions where the curves intersect
coincide with the $\chi^2$ minima of the SED fits, for $A_{V,{\rm
H}\,{II}}= A_{V,{\rm SED}}+1$.
\label{sfrdust.plot}}
\end{figure*}

\subsubsection{Consistency with SED fits}
\label{consist.sec}

We first consider whether the ages, star formation rates, and
extinction derived from the SED fits are consistent with the observed
luminosities and equivalent widths of H$\alpha$. Values implied by the
SED fits were calculated using Eqs.\ \ref{sfr.eq} and \ref{ew.eq}, and
the star formation rates and ages listed in Table~4. Extinction
corrections were applied following the procedure outlined above.

\vbox{
\begin{center}
\leavevmode
\hbox{%
\epsfxsize=6.5cm
\epsffile{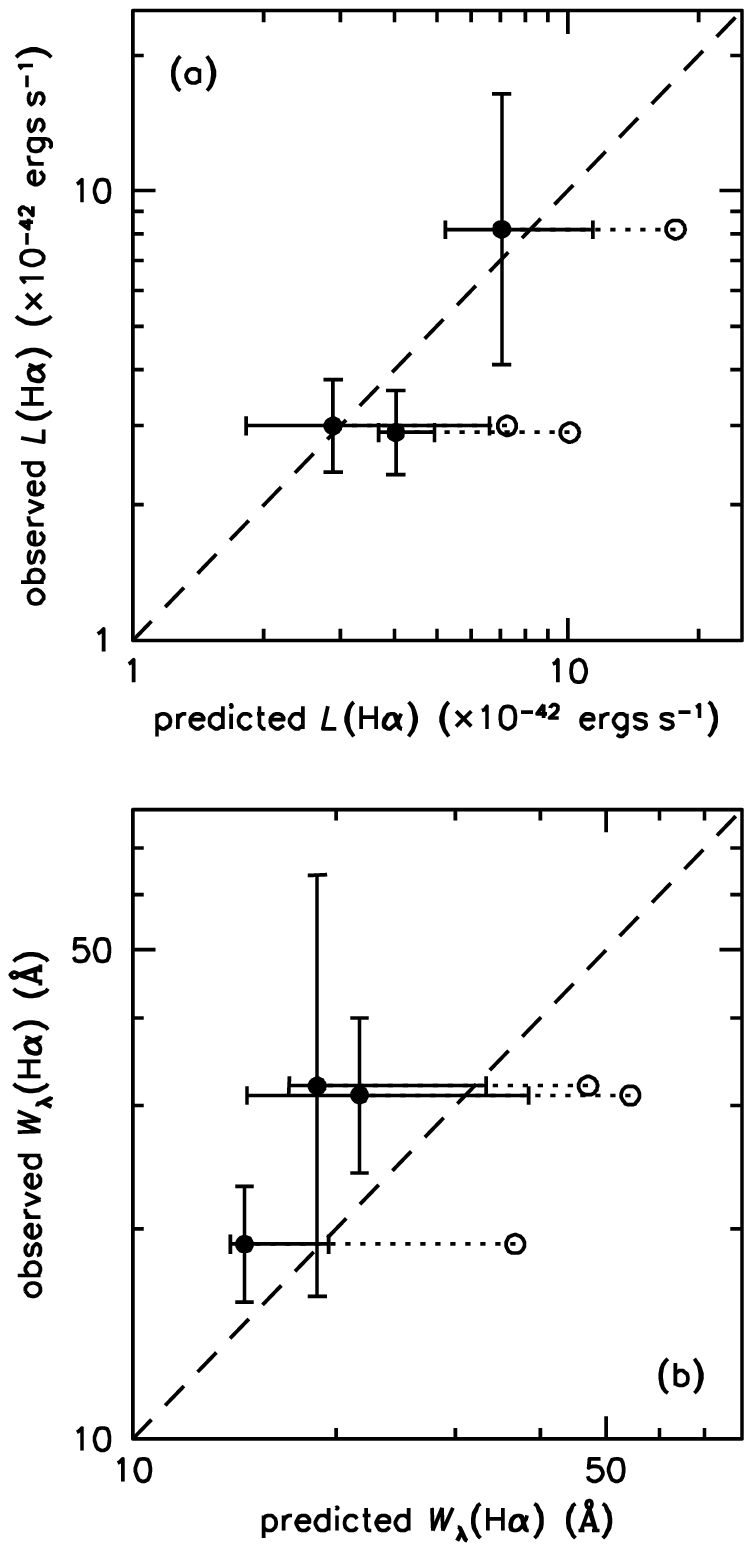}}
\figcaption{\small
Comparison of observed H$\alpha$ line luminosities (a) and
equivalent widths (b) to predicted values from
the SED modeling. Luminosity predictions were computed
from the star formation rates; equivalent widths from the
ages. Open symbols assume
that the extinction toward H{\sc ii} regions is equal to
that of the stellar continuum;
solid symbols assume one magnitude additional extinction.
The H$\alpha$ measurements are in good agreement with the
predictions from the SED modeling.
\label{linecomp.plot}}
\end{center}}

In Fig.\ \ref{linecomp.plot}(a) the line luminosities implied
by the SED fits are
compared to the observed ones. Galaxies
M--1035 and M--1356 are not shown, as their emission lines
are almost certainly dominated by their active nuclei. For C--3396
we use $L_{{\rm H}\alpha} = 2.75
L_{{\rm H}\beta}$ (Osterbrock 1989), after taking differential extinction
into account. 
The SED results overpredict the H$\alpha$ observations for
$A_{V,\,\rm {H\,II}} = A_{V,\,{\rm SED}}$, but they agree very well
if H{\sc ii} regions suffer additional extinction of approximately
one magnitude.

In Fig.\ \ref{linecomp.plot}(b) predicted equivalent widths are
compared to the observed ones. The value for C--3396 was
determined using $W_{\lambda}({\rm H}\beta) =
0.22 W_{\lambda}({\rm H}\alpha)$ ({Jansen} {et~al.} 2000).
Again, the data are in good agreement and
indicate additional extinction of one magnitude
toward H{\sc ii} regions.


\subsubsection{Constraints on $A_V$}

\label{avcon.sec}

Although the agreement between SED fits and H$\alpha$ measurements
is encouraging the two constraints are not truly independent, as
they both rely on the extinction as derived from the broad band
photometry. Furthermore, the additional extinction toward
H{\sc ii} regions is a free parameter in the modeling.
Here we  discuss the dependence
of the age and star formation rate on the dust content in more
detail. We only
consider the two non-active galaxies for which H$\alpha$ has been
reliably measured (M--1383 and M--1319).

We first determine the dependence of the SED fitting results on
the extinction. Rather than having $A_V$ as a free parameter in
the fits it was held fixed at values ranging from $A_V=0$ to $A_V=3$;
only the age and star formation rate were allowed to vary.
As shown in Figure \ref{sfrdust.plot} (blue lines)
the dependencies are quite strong, because the age and star formation rate
are very sensitive to the UV flux, which is a strong function of
$A_V$.

Next we examine the extinction-dependence of the constraints derived from
H$\alpha$ (red lines). The solid lines show
the derived star formation rates and ages assuming $A_{V,\,{\rm H\,II}}
= A_{V,\,{\rm SED}}$. Importantly, the implied ages are higher than
the age of the Universe at $z=2.43$ with this assumption, in particular
for galaxy M--1319. The low H$\alpha$ equivalent widths
thus imply either additional absorption toward H{\sc ii}
regions, or a declining star formation rate (see \S\,\ref{tau.sec}).
Note that this result follows directly from Eq.\,\ref{ew.eq}, and is
independent from the SED fits.
Dashed lines show the effect of an additional magnitude of
extinction for the line emitting gas; with this assumption the
implied ages are lower than the age of the Universe.

Furthermore, the dependencies are much weaker than for the SED fits, because
they are determined by the absorption at $6563$\,\AA\ rather than
in the UV. As an example, for galaxy M--1383
the H$\alpha$-derived star formation rate
varies from $25\,M_{\odot}\,$yr$^{-1}$ to $300\,M_{\odot}\,$yr$^{-1}$
for $A_V=0-3$, whereas the
star formation rate derived from the SED fits varies between
zero and 
$2000\,M_{\odot}\,$yr$^{-1}$.
The different dependencies on the extinction can be utilized:
we can determine $A_V$ by requiring that the ages and star formation
rates derived from the SED fits are equal to those derived
from H$\alpha$. Consistent solutions are obtained at the intersections
of the lines derived from the SED fits and those derived from H$\alpha$.
We find $0.8 \lesssim A_V \lesssim 1.2$ for M--1319 and $1.8 \lesssim
A_V \lesssim 2.6$ for M--1383, if the additional extinction toward
H{\sc ii} regions is approximately one magnitude.

This derivation of the extinction is independent of the values
corresponding to the best fitting SED: there is no ``built-in''
requirement that the solutions with the lowest $\chi^2$
residual should fall in the regions of the
plots where the SED constraints and the H$\alpha$ constraints
are consistent with each other. Nevertheless, they do: the large
solid symbols correspond to the best fitting SEDs (shown in
Fig.\ \ref{seds.plot}); as in Fig.\ \ref{sedresult.plot}
small dots indicate the uncertainty. These results considerably
strengthen the conclusion from \S\,\ref{consist.sec}:
the constraints on ages and star formation rates
from H$\alpha$ are consistent with those derived
from the broad band photometry {\em only} for the values
of $A_V$ corresponding to the best fitting SEDs.
Furthermore, the additional extinction toward H{\sc ii} regions
(which was treated as a free parameter in \S\,\ref{consist.sec})
is required by the strict upper limit on the ages imposed by the age
of the Universe at $z=2.43$.

\subsection{Models with Declining Star Formation Rates}
\label{tau.sec}

The inferred star formation rates
are very sensitive to the (high) extinction that
we derive. We can determine the ``minimum'' star formation rates
for galaxies M--1319 and M--1383 by applying no
extinction correction to their H$\alpha$ luminosities.
The uncorrected star formation rates are 23\,$M_{\odot}$\,yr$^{-1}$
and 24\,$M_{\odot}$\,yr$^{-1}$ respectively, i.e., an order
of magnitude lower than the extinction-corrected values
(see Table~5). As we demonstrated in the preceeding Sections,
the SEDs of the galaxies, and the H$\alpha$ equivalent widths,
cannot be fitted with continuous star
formation models with low extinction. Here we
investigate what values we find for the
extinction and star formation rates
if we allow for the possibility that the star
formation rates are not constant but declining.
We only consider
the two non-active galaxies with measured H$\alpha$
luminosities; a more extensive discussion is given in F\"orster
Schreiber et al.\ (2004), which describes SED fits to the full sample
of galaxies with $K_s<21.7$ and $J_s-K_s>2.3$ in the MS\,1054--03
field.

Galaxies M--1319 and M--1383 were fitted with Bruzual \& Charlot
(2003) models characterized by a star formation history with an
e-folding time $\tau = 300$\,Myr.  Both galaxies are well fitted by
this model; differences in the best fitting $\chi^2$ between the
declining and the continuous model are not statistically significant.
For both galaxies the $\tau$-model gives lower dust content, lower
ages, and lower star formation rates than the continuous model; values
are listed in Table~5. However, the {\em mean} star formation
rate over the lifetime of the galaxy is actually slightly higher in the
$\tau$-model: $507\,M_{\odot}$\,yr$^{-1}$ for M--1383 and
$330\,M_{\odot}$\,yr$^{-1}$ for M--1319. As a result, the
stellar masses in the two models are similar. 
These effects are illustrated in Fig.\ \ref{sfhis.plot}, which
shows the star formation histories of both galaxies for both models.
The effect of the $\tau$-model is to make the star formation history
more concentrated and peaked in time, but it does not substantially
change the total number of stars formed.

We conclude that the extinction and star formation rates are
model-dependent, and can be lower by factors of $2-3$ if the star
formation histories are modified.  However, the stellar masses are
robust, because declining models have a much higher star formation
rate at earlier epochs (see Fig.\ \ref{sfhis.plot}).

\vbox{
\begin{center}
\leavevmode
\hbox{%
\epsfxsize=8.5cm
\epsffile{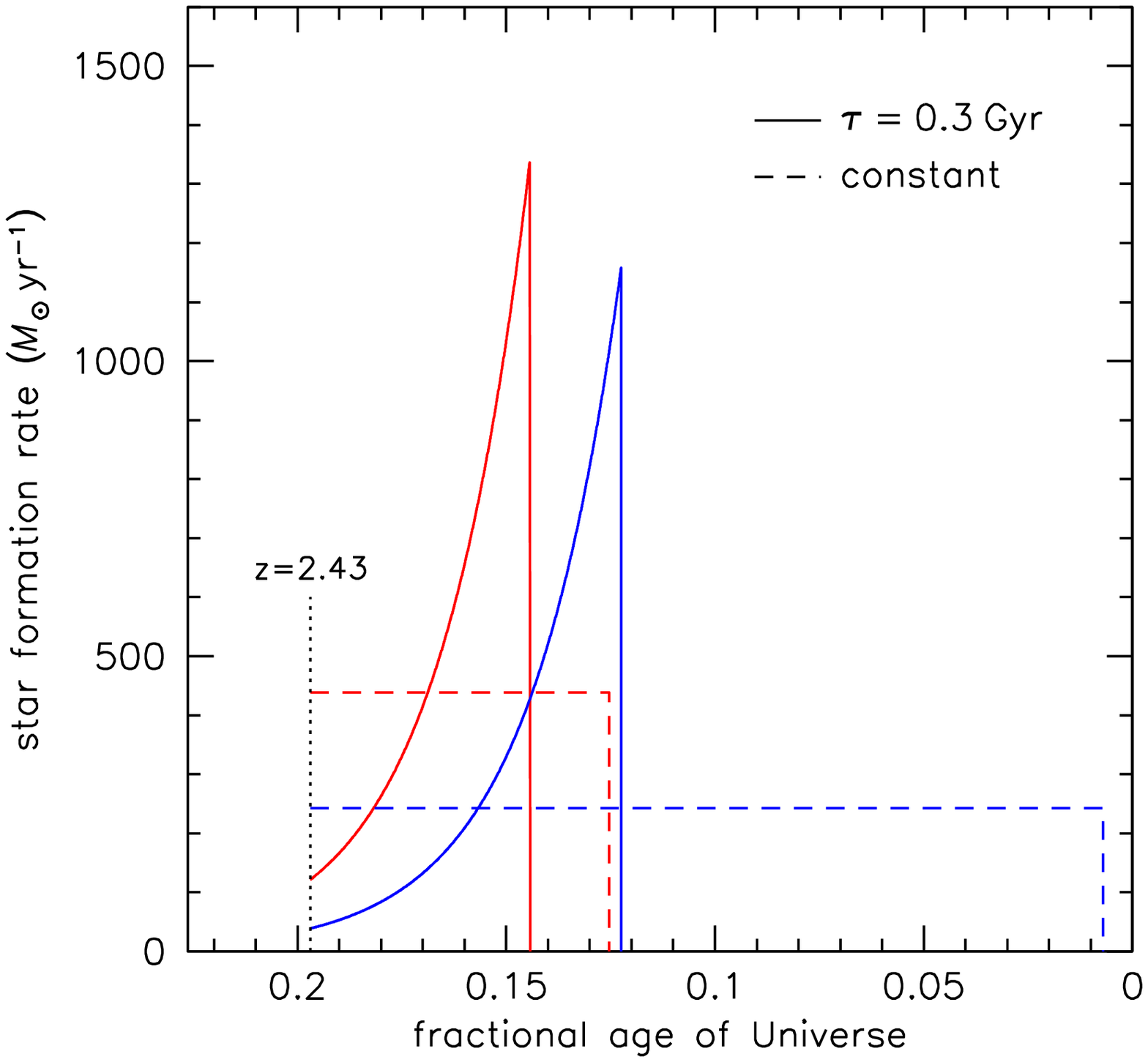}}
\figcaption{\small
Best fitting star formation histories of galaxies M--1383 (red)
and M--1319 (blue), for a continuous star
formation model (dashed) and a $\tau$-model with an e-folding
time of 0.3\,Gyr (solid). Models with declining star formation rates
produce lower star formation rates at the epoch of observation,
but they have much higher star formation rates at earlier epochs.
\label{sfhis.plot}}
\end{center}}

\subsection{Continuum Break}

For one galaxy (M--1319) we obtained a spectrum in the
rest-frame wavelength range 3750\,\AA\ -- 4550\,\AA\
(see \S\,\ref{balmer.sec}), providing
an additional constraint on its stellar population.
The spectrum can be used to disentangle the effects of age
and dust.
Young dusty galaxies have smooth SEDs, whereas galaxies with
stellar populations older than $\sim 500$\,Myr have a
pronounced discontinuity at $\lambda 3625$\,\AA\ (the
Balmer discontinuity) or $\lambda 4000$\,\AA\ (the Ca\,{\sc ii}
H+K break).

The spectrum is of insufficient quality to constrain the age and
dust content independently. Instead, we generated model spectra
for different values of $A_V$ by fitting constant star formation
models to the broad band photometry
(see \S\,\ref{avcon.sec}). This constrained fit ensures that
the best fitting model is consistent with the broad band photometry
(or at least as close as possible for that particular value
of $A_V$). The models were binned
in the same way as the observed spectrum, and fitted using
a $\chi^2$ minimalization.

The best fitting $A_V = 0.6^{+1.2}_{-0.6}$, consistent with
the extinction derived from the broad band photometry ($A_V=1.3 \pm 0.1$)
and from the comparison
of SED fits and H$\alpha$ constraints ($A_V=0.8-1.2$). The uncertainties are
quite large, partly because
the spectrum does not extend blueward of the Balmer break.
Models with extinction higher than $A_V \approx 2$ fit poorly because
they overpredict the flux blueward of 3900\,\AA\ and underpredict the
flux redward of 4400\,\AA; an example with $A_V=3$ is shown by
the red histogram in Fig.\ \ref{balmer.plot}. The best fit to the
broad band photometry is shown in blue; this model is a good match
to the observed spectrum.


\subsection{SCUBA Detection}

Independent constraints on the star formation rate are very important
as they can help break the degeneracy between the form of the
star formation history and the current star formation rate
(see \S\,\ref{tau.sec}).
K.\ Knudsen et al.\ (in preparation)
obtained a deep, large area SCUBA map
of the \msclus\ field, and we have performed an initial investigation
of coincidences of SCUBA detections and $J_s-K_s$ selected galaxies.
Interestingly galaxy M--1383, the object in our sample with the
highest inferred dust content, is very likely a SCUBA source. 

The object is detected with an 850\,$\mu$ flux density of $5.1 \pm 1.5$\,mJy.
Assuming the infrared SED is similar to that of Arp\,220 the
IR luminosity is $5 \times 10^{12}\,L_{\odot}$, and the implied
star formation rate is $\sim 500\,M_{\odot}$\,yr$^{-1}$, in excellent agreement
with the (extinction-corrected) value of $\sim 400\,M_{\odot}$\,yr$^{-1}$
that we determine from the stellar SED 
and H$\alpha$ in the constant star formation model (see Table 5).
We note that such a good agreement is uncharacteristic of
local ultra-luminous infrared galaxies, as star forming regions in
these galaxies usually suffer from  such high visual extinction that it is very
difficult or impossible to correct for it.
A full analysis of submm and X-ray constraints on the star formation rates
of \orgs\ will be presented elsewhere; here we note that the submm
data for this galaxy
provide strong independent support for our modeling procedures.

\subsection{Metallicities}
\label{Z.sec}

Determining metallicities for high redshift galaxies is
difficult, as generally only a limited number of (bright) emission
lines is available. The most widely used metallicity indicator
for faint galaxies is the $R_{23}$ index, which relates O/H
to the relative intensities of [O\,{\sc ii}] $\lambda 3727$,
[O\,{\sc iii}] $\lambda\lambda 4959,5007$, and H$\beta$
({Pagel} {et~al.} 1979; {Kobulnicky}, {Kennicutt}, \&  {Pizagno} 1999). Using this indicator {Pettini} {et~al.} (2001)
derive abundances for five LBGs of
between $0.1\times$ and $1\times$ Solar, with the main
uncertainty the double-valued nature of the $R_{23}$ index.

We do not have the required measurements to calculate the
$R_{23}$ index for any of the four non-active galaxies
in our sample. However, for two objects the luminosities of
H$\alpha$ and [N\,{\sc ii}] $\lambda 6584$ are available,
enabling us to derive metallicities using the ``N2''
index. This indicator is defined as N2$ =
\log$\,([N\,{\sc ii}]\,\,$\lambda 6584 /$\,H$\alpha$); it
was first suggested by {Storchi-Bergmann}, {Calzetti}, \&  {Kinney} (1994) and discussed
recently by {Denicol{\' o}}, {Terlevich}, \&  {Terlevich} (2002). Advantages of the N2
indicator are that it has a single-valued dependence on the
oxygen abundance and that it is insensitive to reddening.
However, it is sensitive to ionization and,
as oxygen is not measured directly, to O/N variations
(see, e.g., {Denicol{\' o}} {et~al.} 2002).

For objects M--1383 and M--1319 we find ${\rm N}2 = -0.40 \pm 0.16$
and ${\rm N}2 = -0.10 \pm 0.13$ respectively. Using the calibration
of {Denicol{\' o}} {et~al.} (2002),
\begin{equation}
12 + \log({\rm O/H}) = 9.12 (\pm 0.05)+0.73(\pm 0.10) \times {\rm N2},
\end{equation}
we find $12+\log({\rm O/H}) = 8.8 \pm 0.3$ and $9.0\pm 0.3$ respectively
(where we used an intrinsic scatter in the method of 0.2 dex; see
Denicol\'o et al.\ 2002).
As the Solar abundance $12+\log({\rm O/H})_{\odot}=8.83$, we infer that
the abundances of the two $J_s-K_s$ selected galaxies are $1\times$
and $1.5\times$ Solar, with uncertainties of a factor of two.
We conclude that current evidence suggests that
red $z>2$ galaxies are already
quite metal rich, as might have been expected from their high
dust content and extended star formation histories.

\section{Kinematics}

\subsection{Masses and $M/L$ Ratios}
\label{mass.sec}

The stellar masses implied by the SED fits are substantial, ranging
from $1.3-4.9 \times 10^{11}\,M_{\odot}$,
with median $3.5\times 10^{11}\,M_{\odot}$.  Although these results
are supported by the H$\alpha$ luminosities and equivalent widths
significant systematic uncertainties remain:
the SED fitting is dependent on the assumed IMF and metallicity, and
both methods are sensitive to the assumed star formation history
and extinction law.

Independent constraints on the masses of the red galaxies can be
obtained from their kinematics. Line widths were measured for four
galaxies, one of which harbors an AGN (see \S\,\ref{width.sec}).  The
widths are substantial, and range from 134\,\kms\ to 240\,\kms.  The
line broadening can be caused by regular rotation in a disk, random
motions, or non-gravitational effects. {Rix} {et~al.} (1997) derive an
empirically motivated transformation from observed line width to
rotation velocity in a disk; applying their relation $\sigma \approx
0.6 V_c$ gives rotation velocities in the range 200--400\,\kms.
The kinematics are thus in qualitative agreement with the high
masses inferred for the stars.

The large linewidths imply that the
four red galaxies are very massive,
unless they are highly compact.
The galaxies are faint in the rest-frame UV which makes it difficult
to measure their sizes from optical HST images. Fig.\ \ref{size.plot}
shows $K_s$ images of the four galaxies with
measured velocity dispersions, all taken with ISAAC on the VLT
in $\approx 0\farcs 5$ seeing.
All four galaxies are resolved in these ground-based images.
Effective radii are $\sim 0\farcs 5$, with the largest galaxy (M--1383)
having $r_e \approx 0\farcs 7$ (Trujillo et al.\ 2004 and
in preparation).
Using $r_e \approx 0\farcs 5$ as a reasonable estimate for the size,
the median $\sigma \approx 200$\,\kms\ for the velocity dispersion,
and the same normalization as {Pettini} {et~al.} (2001), we find
$M_{\rm dyn} \sim 2 \times 10^{11}\,M_{\odot}$. This number
is in good agreement with the stellar mass estimates.

\vbox{
\begin{center}
\leavevmode
\hbox{%
\epsfxsize=8cm
\epsffile{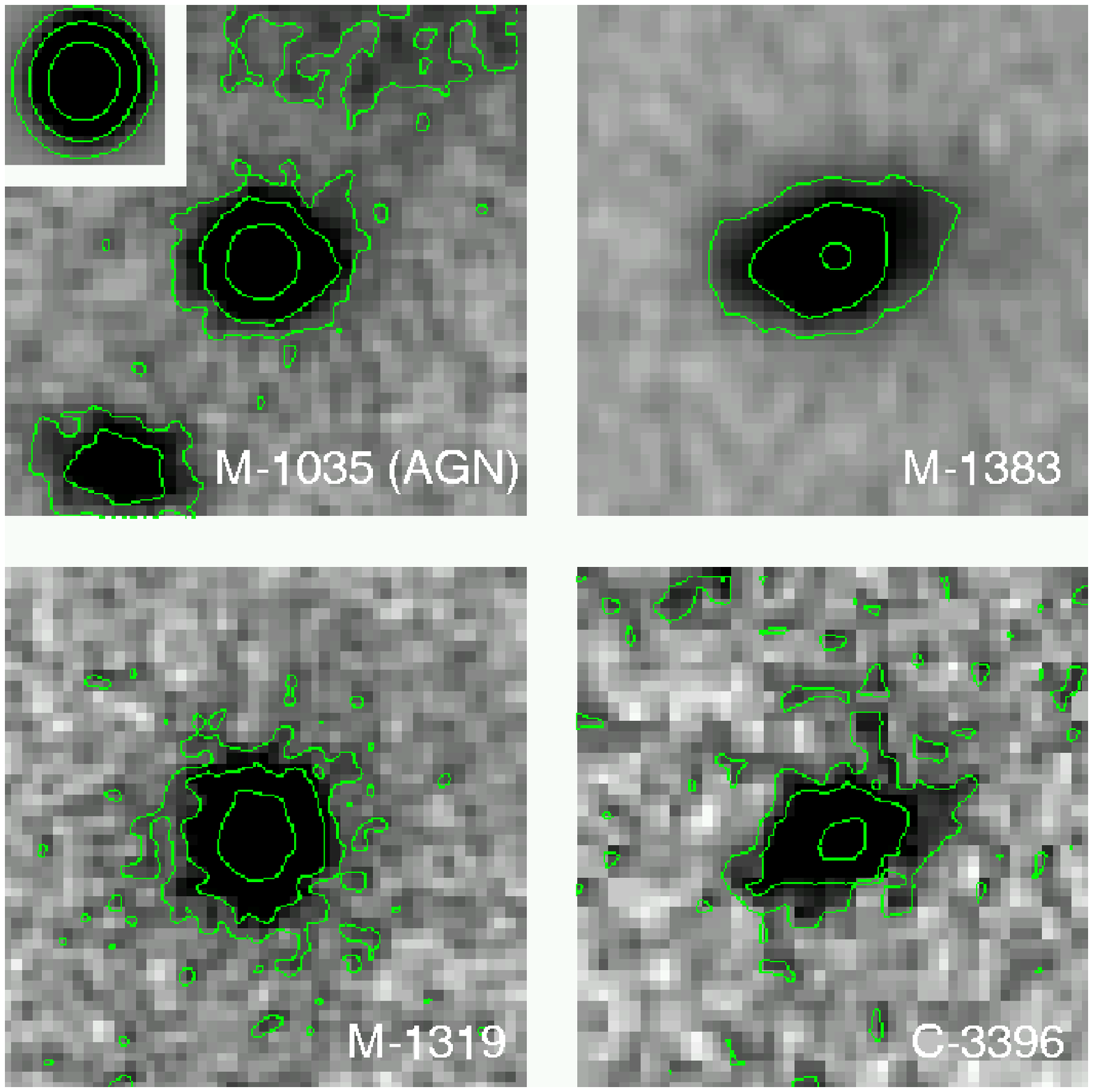}}
\figcaption{\small
Ground based $K_s$ images of the four galaxies with measured
line widths, obtained with ISAAC on the VLT. Images are
$5'' \times 5''$; the FWHM image quality is $\approx 0\farcs 5$.
All four galaxies are resolved; they have typical effective
radii of $\sim 0\farcs 5$.
\label{size.plot}}
\end{center}}

\begin{figure*}[t]
\epsfxsize=17.5cm
\epsffile{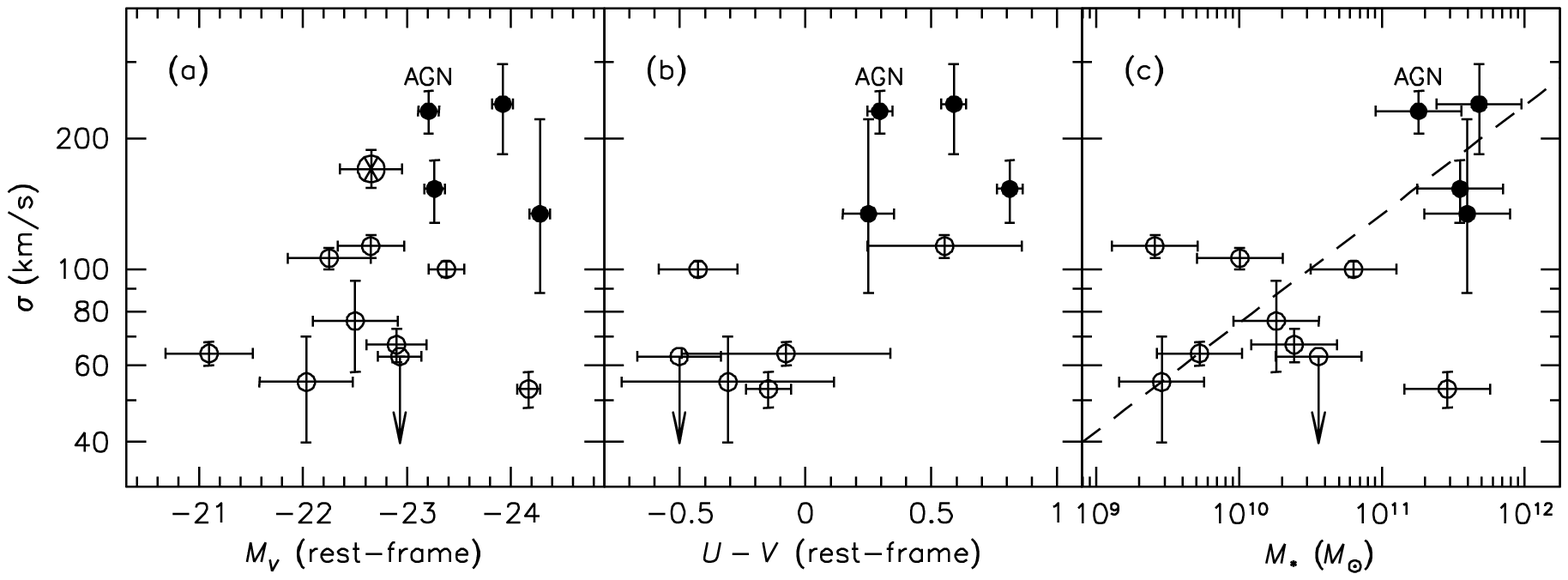}
\caption{\small
Correlations of rest-frame optical linewidth with rest-frame $V$-band absolute
magnitude (a), rest-frame $U-V$ color (b), and stellar mass
(c). Small open circles are Lyman-break galaxies at $z\approx 3$, the
large open symbol is a LBG at $z=2.19$, and filled circles
are $J_s-K_s$ selected galaxies with mean $z=2.6$. Linewidth correlates with
luminosity, color, and stellar mass in the combined sample.
The line in (c) is the relation between rotation velocity and baryonic
mass determined for local galaxies by McGaugh et al.\ (2002),
using $\sigma \approx 0.6 V_c$ (Rix et al.\ 1997). This relation provides
an excellent fit to the $z> 2$ galaxies.
\label{sigcorr.plot}}
\end{figure*}

The median absolute $V$ magnitude of the four galaxies is $-23.6$ (see
below). With $M \sim 2 \times 10^{11}\,M_{\odot}$ this implies
dynamical mass-to-light ratios $M/L_V \sim 0.8$ in Solar units, a factor of $\sim
5$ higher than $z\approx 3$
Lyman-break galaxies ({Pettini} {et~al.} 2001). Based on the range of
values within our sample we estimate that the uncertainty in this number is
approximately a factor of two.


\subsection{Correlations with Linewidth}

{Pettini} {et~al.} (2001) studied Tully-Fisher ({Tully} \& {Fisher} 1977)
like correlations for LBGs, by comparing their
kinematics (as measured by the line width of rest-frame optical
emission lines) to their rest-frame UV and optical magnitudes. 
No correlations were found over the range $50$\,\kms\,$\leq\sigma
\leq$\,115\,\kms. Here we add the four red $z>2$ galaxies
with measured linewidths to Pettini's sample and re-assess
the existence of correlations with linewidth at $z\sim 3$.

Rest-frame $V$ magnitudes
were determined from observed $K_s$ magnitudes
and $J_s-K_s$ colors.
For each galaxy we derive a transformation
between observed magnitudes and rest-frame magnitudes of the
form
\begin{equation}
\label{trafo.eq}
V_{\rm rest} = K_s + \alpha (J_s-K_s) + \beta.
\end{equation}
The constants $\alpha$ and $\beta$ depend on the redshift, but
only weakly on the spectral energy distribution of the galaxy
(see {van Dokkum} \& {Franx} 1996). Using spectral energy distributions
from {Bolzonella} {et~al.} (2000) we determined that, for $z\sim 3$, the transformations
have the same form within $\sim 5$\,\% for SEDs ranging from
starburst to early-type. {Shapley} {et~al.} (2001) give
$K_s$-band magnitudes for nine galaxies in the
{Pettini} {et~al.} (2001) sample, and $J-K_s$ colors for six.
For the three LBGs that have not
been observed in $J$ we use synthetic $J-K_s$ colors;
the added uncertainty in $V_{\rm rest}$
is negligible as $\alpha=-0.01$ for $z=3$.

Figure\ \ref{sigcorr.plot}(a) shows the ``Tully-Fisher''
diagram of rest-frame optical linewidth (corrected for
instrumental resolution) versus rest-frame absolute
$V$ magnitude, for $z>2$ galaxies. The four $J_s-K_s$ selected galaxies
have higher linewidths and are on average more luminous than $z\approx 3$
LBGs. Although visually suggestive,
the Spearman rank and Pearson correlation tests give
probabilities of $\sim 50$\,\%
that $\log \sigma$ and $M_V$ are uncorrelated, and we conclude
the evidence for the existence of a luminosity-linewidth relation at
$z\sim 3$ is not yet conclusive. {Erb} {et~al.} (2003) find that
$z\approx 2$ LBGs have on average higher linewidths than those
that have been observed by {Pettini} {et~al.} (2001)
at $z\approx 3$, and it will
be interesting to see where these galaxies lie on the Tully-Fisher
diagram. The only $z\approx 2$ LBG from {Erb} {et~al.} (2003) with measured $K$
magnitude (Q1700-BX691 at $z=2.19$) is indicated with a large
open symbol in Fig.\ \ref{sigcorr.plot}(a).

In Fig.\ \ref{sigcorr.plot}(b) linewidth is plotted versus
rest-frame $U-V$ color, which were derived from the observed $J_s - K_s$
colors in a similar way as the rest-frame $V$ magnitudes.
Only the six Lyman-break galaxies
with measured $J-K_s$ colors are included. There is a striking correlation,
with significance $>97$\,\%. The figure also clearly demonstrates
the large difference in rest-frame colors between LBGs and
$J_s-K_s$ selected galaxies. In the local universe similar
trends exist, with more massive galaxies generally being redder.

Finally, Fig.\ \ref{sigcorr.plot}(c) shows the relation between
linewidth and stellar mass. To limit systematic effects
we determined the stellar masses of the LBGs with our own methods,
using the photometry presented in
{Shapley} {et~al.} (2001) and {Steidel} {et~al.} (2003).
The median stellar mass is $2 \times 10^{10}\,M_{\odot}$,
very similar to what {Shapley} {et~al.} (2001) finds for the
same nine objects. 
The correlation has a significance of
$\approx 90$\,\%. The line shows the ``baryonic Tully-Fisher
relation'' of McGaugh et al.\ (2000), which is an empirical
relation between rotation velocity and total mass in stars and
gas (for galaxies with $M_*>10^9\,M_{\odot}$ this
is approximately equal to the stellar mass alone).
With $\sigma \approx 0.6 V_c$ this relation
takes the form $M_* = 312\,\sigma^{4}\,M_{\odot}$ in our cosmology; without
any further scaling it provides
a remarkably good fit to the $z>2$ data.

\begin{figure*}[b]
\epsfxsize=16cm
\epsffile[-60 194 553 646]{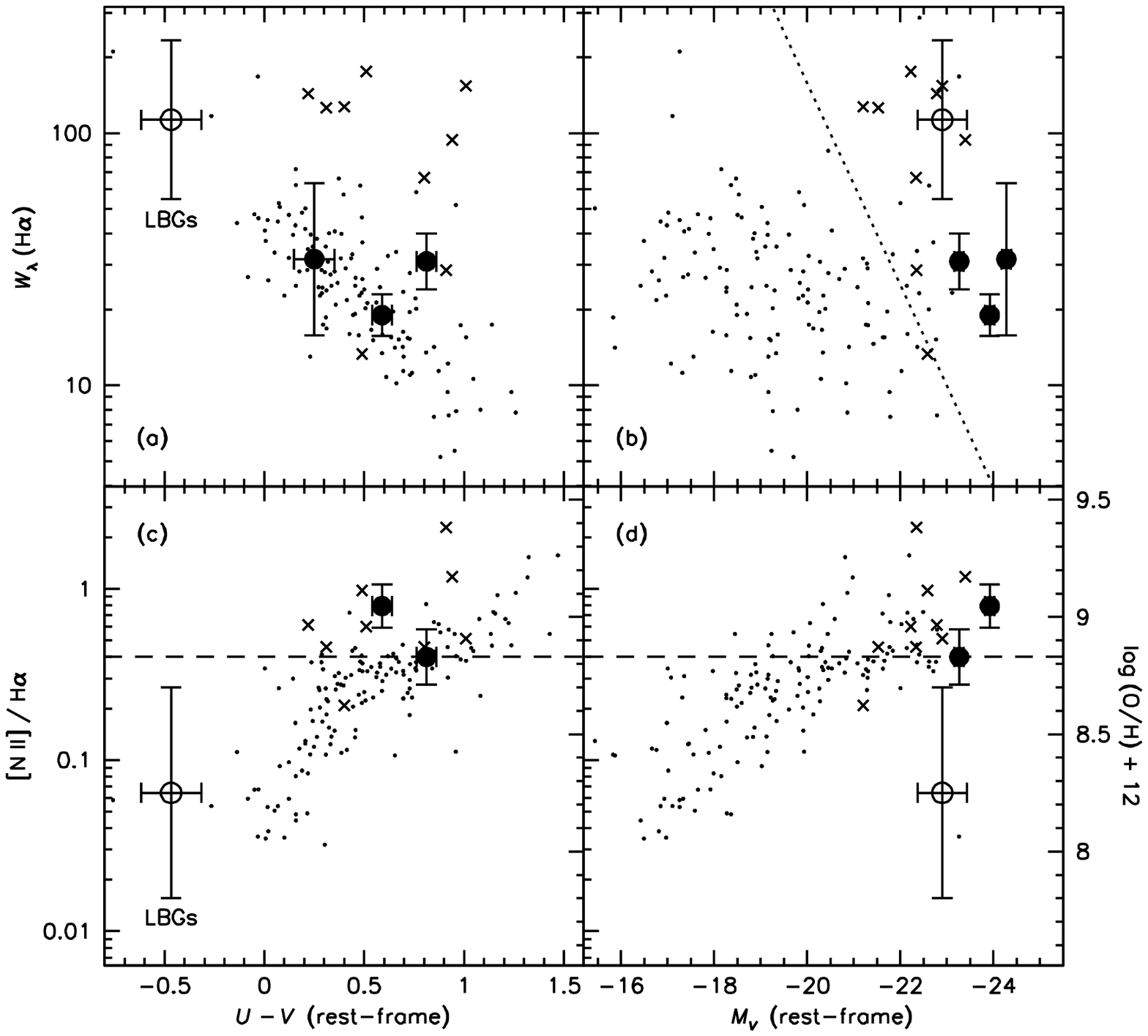}
\caption{\small
Correlations of H$\alpha$ equivalent width (a,b) and the
[N\,{\sc ii}]\,\,$\lambda 6584 /$\,H$\alpha$ ratio (c,d) with rest-frame
$U-V$ color and $V$-band absolute magnitude, for samples of
star-forming galaxies.
Shown are nearby normal galaxies from the Nearby Field Galaxy
Survey (Jansen et al.\
2000,2001) (small dots), nearby Luminous InfraRed Galaxies
from Arhus et al.\ (1989) (crosses), 
LBGs from Pettini et al.\ (2001) (open circles), and the three
non-active \orgs\ with measured Balmer lines from the present study
(large solid circles). The metallicity scale on the right axis
of (c) and (d) corresponds to the LBGs from Pettini et al.\ (2001). This
scale is coupled to the [N\,{\sc ii}]\,/\,H$\alpha$ scale through the
Denicolo et al.\ (2002) calibration.  No extinction corrections
were made. The broken line in (b)
indicates a line of constant H$\alpha$ luminosity; the dashed
line in (c) and (d) indicates Solar metallicity.
Contrary to LBGs,
the red $z>2$ galaxies have similar colors, metallicities, and
H$\alpha$ equivalent widhts as normal nearby galaxies.
\label{colcorr.plot}}
\end{figure*}

We note that the analysis may be affected by selection effects,
as we combine very different samples of galaxies
in Fig.\ \ref{sigcorr.plot}. Pettini's LBGs are at slightly higher
mean redshift than the \orgs, and their linewidths were measured from
the [O\,{\sc iii}] lines instead of H$\alpha$. The $z\approx 2$
LBG sample of {Erb} {et~al.} (2003) shows a much larger range in linewidth
than Pettini's $z\approx 3$ sample.
This may be caused by differences in selection (the Pettini sample
was selected to be very bright in the rest-frame UV, which may select
relatively low mass objects; see Erb et al.\ 2003), but may also
hint at systematic differences between linewidths determined from
H$\alpha$ and the oxygen lines, or even evolution.
It will be interesting to see where the Erb et al.\ objects fall
in the panels of Fig.\ \ref{sigcorr.plot}, especially those
with the highest linewidths.

We are also limited by small number statistics.
In particular, the reason why linewidth appears to correlate
most strongly with color may simply be that there is no
$J$-band measurement given by {Shapley} {et~al.} (2001)
for West MMD11,
which is a luminous and apparently massive Scuba source
with very small linewidth ($\sigma = 53 \pm 5$\,\kms).
When this object is arbitrarily removed in Figs.\ \ref{sigcorr.plot}(a)
and (c) as well the correlation significance jumps to
$\approx 98$\,\% in both cases.

\subsection{Outflows}

A major result of kinematic studies of LBGs is that they exhibit
galactic-scale outflows 
({Franx} {et~al.} 1997; {Pettini} {et~al.} 1998, 2000; {Adelberger} {et~al.} 2003). The main evidence comes
from comparisons of the radial velocities of Ly$\alpha$, UV absorption
lines, and rest-frame optical emission lines.  The UV absorption lines
are blueshifted with respect to the optical lines as they trace the
gas flowing towards us, whereas Ly$\alpha$ emission appears redshifted
as it originates from the back of the shell of outflowing material
(see, e.g., Pettini et al.\ 2001 for details).
In LBGs winds have typical velocities of $\sim 300$\,\kms\ (with
a large range), and are thought to play an important role
in enriching their surrounding medium ({Adelberger} {et~al.} 2003).

In our sample of
\orgs\ galactic winds may be less efficient in
enriching the IGM; as the galaxies have
very high masses the mechanical energy provided by
supernovae and stellar winds may not be sufficient to let the
gas escape. We can constrain the wind velocities for only
two galaxies in our sample: C--3396, for which we have measured
Ly$\alpha$, UV absorption lines, and the \oiii\ lines, and
M--1319, with Ly$\alpha$ and H$\alpha$. In C--3396 the
implied wind velocity is $\sim 300$\,\kms, similar to typical values for
LBGs. In M--1319 it is only $\sim 100$\,\kms. Given the large
range observed for LBGs, much larger samples are needed to reliably
compare the wind velocities of \orgs\ and LBGs, and to determine
whether the wind velocities of \orgs\ exceed their escape
velocities.

\section{Comparison to Other Galaxies}

The $J_s-K_s$ selection technique is one among many ways to select
distant galaxies. Examples of other galaxy populations beyond $z\sim
2$ are Lyman break galaxies ({Steidel} {et~al.} 1996), Ly$\alpha$ emitters
({Hu}, {Cowie}, \& {McMahon} 1998), hard X-ray sources (e.g., {Barger} {et~al.} 2001), and
submm sources (see, e.g., {Blain} {et~al.} 2002).  With the growing
diversity of the high redshift ``zoo'' it becomes increasingly
important to determine the relations between these populations, with
the aim of establishing plausible evolutionary histories for nearby
galaxies.

\subsection{Comparison to Star Forming Galaxies at $z=0-3$}

Star formation in local and distant galaxies usually takes place
in one of two modes: low intensity, prolonged star formation
in a disk (most nearby galaxies), or high intensity, relatively short
lived star formation (star burst galaxies, LIRGs and ULIRGs, and
probably also Lyman break galaxies). The \orgs\ studied in this
paper seem to have different star formation histories: high
intensity star formation that is maintained for a long
time (up to several Gyr).

These trends are illustrated in Fig.\ \ref{colcorr.plot}, which
shows correlations between emission line properties, color,
and absolute magnitude for \orgs, LBGs, normal nearby galaxies
from Jansen et al.\ (2000), and Luminous InfraRed Galaxies
(LIRGS) from {Armus}, {Heckman}, \& {Miley} (1989). In (a) we
show the relation between H$\alpha$ equivalent width and rest-frame
$U-V$ color. For C--3396 and the LBGs the equivalent widths
were determined using the empirical relation
$W_{\lambda}({\rm H}\beta) =
0.22 W_{\lambda}({\rm H}\alpha)$
({Jansen} {et~al.} 2000).
The equivalent width is a measure of the ratio of present
to past star formation, and therefore a good indicator of the time
that has elapsed since the onset of star formation. Lyman break
galaxies have high values of $W_{\lambda}({\rm H}\alpha)$, similar to
the much redder LIRGS, consistent with the
burst-like, short-lived nature of their star formation.  The colors
of \orgs\ are similar to those of LIRGS, but their equivalent widths
are similar to those of normal nearby
galaxies, indicating prolonged star formation histories.
As can be seen in Fig.\ \ref{colcorr.plot}(b) \orgs,
LBGs, and LIRGS all have very high luminosities compared to
normal galaxies. The slope of the dotted
line indicates constant H$\alpha$ luminosity;
the H$\alpha$ luminosities for the samples of
LBGs, \orgs, and LIRGS are similar.

In Fig.\ \ref{colcorr.plot}(c) we show the relation between
the [N\,{\sc ii}]\,/\,H$\alpha$ ratio and color.
Local galaxies show a strong relation in this diagram,
a reflection of the fact that more
metal rich galaxies are redder.
The two \orgs\ fall on the relation defined by local
galaxies, and have similar or slightly lower metallicities
than LIRGs. The \orgs\ also fall on the local relation
of  [N\,{\sc ii}]\,/\,H$\alpha$ ratio and luminosity (Fig.\ \ref{colcorr.plot}d).
For LBGs the [N\,{\sc ii}]\,/\,H$\alpha$ index has not been measured
due to the fact that H$\alpha$ falls outside the $K$
window. The open circles in (c) and (d) were taken from Pettini et al.\ (2003),
and correspond to the right axis. The relation between
the left and right axes is that of
{Denicol{\' o}} {et~al.} (2002). The two \orgs\ have higher metallicities than
most LBGs, consistent with their redder colors and older ages.

We conclude that the \orgs\ fall on the local relations between
H$\alpha$ equivalent width
and color, metallicity and color,
and metallicity and luminosity. They have similar colors as
LIRGS but lower $W_{\lambda}({\rm H}\alpha)$, indicating more
prolonged star formation. LBGs do not fall on the relations
defined by normal nearby galaxies, as they have
a low metallicity for their luminosity and are much
bluer (see also, e.g., {Pettini} {et~al.} 2001).

\subsection{Comparison to Early-type Galaxies}

It seems difficult to avoid the conclusion that the seven
galaxies studied here are very massive. Three lines of evidence
imply masses $\gtrsim 10^{11}\,M_{\odot}$: stellar population
synthesis model fits  to the
$UBVRIJHK$ spectral energy distributions (all seven
galaxies); the luminosity and
equivalent width of H$\alpha$ (two galaxies); and
the line widths combined with approximate sizes (four galaxies).

Given the high masses, red rest-frame optical colors, and apparent
clustering ({Daddi} {et~al.} 2003; {van Dokkum} {et~al.} 2003) of $J_s-K_s$ selected galaxies we can
speculate that some or all of their descendants are early-type
galaxies. It is therefore
interesting to compare the masses and $M/L$ ratios to those of
early-type galaxies at lower redshift.  Such comparisons are fraught
with uncertainties, and given the small size of our sample should
be regarded as tentative explorations.

In Fig.\ \ref{sdss.plot} we compare the stellar masses (derived from the
SEDs) of
the seven galaxies in our sample to the stellar mass distribution
of  elliptical\footnote{Their
definition of elliptical galaxies includes many S0 galaxies.} galaxies in
the Sloan Digital Sky Survey (SDSS), as compiled by
Padmanabhan et al.\ (2003). The SDSS sample includes both field
and cluster galaxies. The distribution of
the $J_s-K_s$ selected galaxies is similar to that of
local elliptical galaxies. In particular, if we require that their
present-day
masses should not exceed those of the most massive
galaxies observed in the local
universe, their stellar masses cannot grow by
much more than a factor of 2--3 after
$\langle z \rangle = 2.6$.

\vbox{
\begin{center}
\leavevmode
\hbox{%
\epsfxsize=8cm
\epsffile{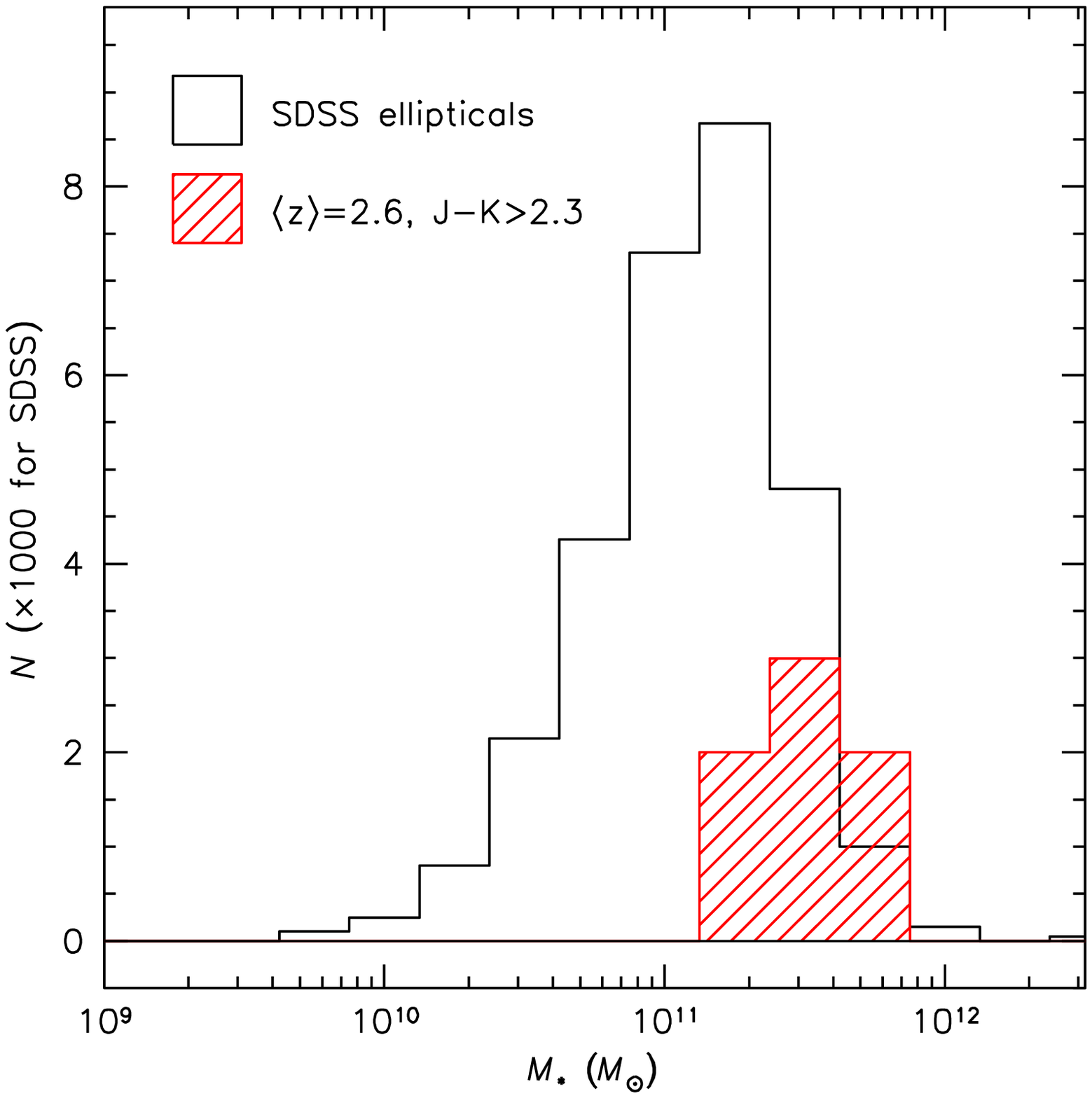}}
\figcaption{\small
Comparison of the stellar masses (derived from SEDs) of the seven $J_s-K_s$ selected
galaxies discussed in the present paper to those of $>29,000$
elliptical
galaxies in the Sloan Digital Sky Survey (Patmanabhan et al.\ 2003).
The red $z>2$ galaxies have similar stellar masses as
massive elliptical galaxies in the nearby universe.
\label{sdss.plot}}
\end{center}}


Comparisons of luminosities or masses are very sensitive to selection
effects. These effects are less pronounced when comparing ratios of
mass and luminosity.  In Fig.\ \ref{mlevo.plot} we compare the dynamical
$M/L$ ratios of the red $z>2$ galaxies to those of nearby early-type
galaxies (Faber et al.\ 1989; J\"orgensen et al.\ 1996), early-type
galaxies at $0<z<1.3$ ({van Dokkum} \& {Stanford} 2003; {van Dokkum} \& {Ellis} 2003), and Lyman-break galaxies
({Pettini} {et~al.} 2001).  As most of the previous work was done in
rest-frame $B$, we transformed the rest-frame $M/L_V$ ratios of the
$J_s-K_s$ selected galaxies to $M/L_B$ using the observed $J_s-K_s$
colors (see Eq.\ \ref{trafo.eq}); we find $M/L_B \approx M/L_V \approx
0.8$ in Solar units.  Based on the range of $M/L$ ratios within our
sample we estimate that the uncertainty in this number is
approximately a factor of two.  The high redshift data were tied to the
work on early-type galaxies in the following way. For the cluster
MS\,1054--03 we determined rest-frame $M/L$ ratios and masses in the
same way as for the $z>2$ galaxies, using their effective radii,
velocity dispersions, and total rest-frame $B$ magnitudes.  The $M/L$
ratios correlate with the mass. We fitted a powerlaw to this
correlation, and determined the $M/L_B$ ratio corresponding to the
median mass of the $z>2$ \orgs.  The MS\,1054--03 data were tied to
the other cluster- and field galaxies using the offsets derived from
the Fundamental Plane (see, e.g., {van Dokkum} {et~al.} 2001).
We note that this procedure gives $M/L$ ratios at $z=0$ 
which are similar to the
average $M/L$ ratios measured for nearby elliptical galaxies:
{van der Marel} (1991) finds $M/L_B \approx 7 (M/L)_{\odot}$, and
our method gives $M/L_B \approx 8 (M/L)_{\odot}$ at $z=0$.

\vbox{
\begin{center}
\leavevmode
\hbox{%
\epsfxsize=8cm
\epsffile{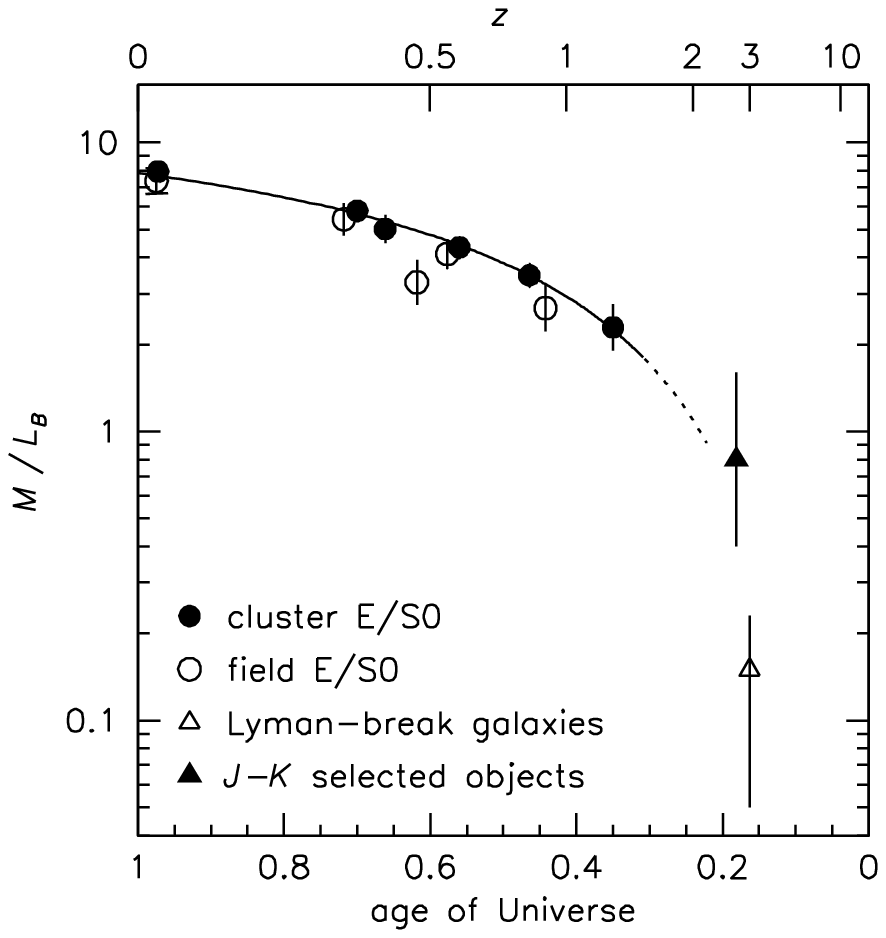}}
\figcaption{\small
Dynamical mass-to-light ratios of early-type galaxies at $0<z<1.3$
and galaxies at $z>2$. The solid points probably represent
the most massive galaxies in the universe at any epoch.
The line is a fit
to the redshift evolution of cluster galaxies, and has the form
$\log M/L_B = 0.89-0.43 z$. The extrapolation to
higher redshifts is (obviously) still very uncertain.
\label{mlevo.plot}}
\end{center}}

The $M/L$ ratios of the red $z>2$ galaxies are lower by a factor of $\sim 3$
than those of the highest redshift early-type galaxies yet measured.
They are consistent with a simple empirical extrapolation: the
line is a fit to the cluster galaxies of the form
\begin{equation}
\label{mlevo.eq}
\log M/L_B = 0.89-0.43 z.
\end{equation}
It is tempting to interpret this figure in terms of simple passive
evolution. However, progenitor bias ({van Dokkum} \& {Franx} 2001) and other
selection effects would undermine the value of such an analysis,
at least beyond $z=1.3$.
Furthermore, the models presented here suggest that the \orgs\
are not passively evolving, but still forming new stars.
Finally, dust
corrections have not been applied.
Rather than interpreting the figure in terms of stellar evolution,
the solid points can be viewed as the {\em observed} change in
mass-to-light ratio of massive galaxies as a function of redshift.
Galaxies with $M/L$ ratios higher than the broken line may be
very rare, and the empirical Eq.\ \ref{mlevo.eq} can provide feducial upper
limits to the masses of distant galaxies when only their luminosities
are known.

\section{Conclusions}

Near-infrared spectroscopy with Keck/NIRSPEC has provided new
constraints on the nature of red galaxies at redshifts $z>2$.  The results
are not unique, as they depend on the assumed functional form of star
formation history, and on the (unknown) extinction towards the line
emitting gas.  The low H$\alpha$ equivalent widths imply either
significant extinction or a declining star formation rate. For
constant star formation models we find total extinction toward H{\sc ii}
regions of $A_V = 2-3$\,mag, extinction-corrected star formation rates
of 200--400 $M_{\odot}$\,yr$^{-1}$, ages $1-2.5$\,Gyr, and stellar
masses $1-5\times 10^{11}\,M_{\odot}$. In declining models the star
formation rates and extinction are lower but the stellar masses are
similar. Measurements of H$\alpha$ and H$\beta$ for the same objects
will help constrain the extinction toward the
line-emitting gas.

The $J_s-K_s$ selected objects extend
the known range of properties of early galaxies to
include higher masses, mass-to-light ratios, and ages allowing
a re-evaluation of the existence of galaxy scaling relations in the
early universe. By combining our small sample of \orgs\ with a published
sample of LBGs we find significant correlations between
line width and color and line width and stellar mass. The latter
correlation is intruiging, as it has the same form as the ``baryonic
Tully-Fisher relation'' in the local universe. There are, however,
many uncertainties.  In particular, it is unclear whether the line
widths at $z>2$ measure regular rotation in disks, and if they do,
whether they sample the rotation curves out to the same radii as in
local galaxies. Also, the correlations are currently only marginally
significant, and larger samples are needed to confirm them.
Finally, the comparison to
$z\approx 3$ LBGs may not be appropriate: initial results from
studies at $z\approx 2$ indicate that their measured properties
are markedly different from the $z\approx 3$ LBGs, with
the lower redshift galaxies being more massive and more metal
rich (Erb et al.\ 2003 and private communication). A plausible
reason is that the $z\approx 3$ LBGs were selected to be bright
in the rest-frame ultraviolet, which may select lower
mass and lower metallicity objects (Erb et al.\ 2003). It will
be interesting to see if the $z\approx 2$ LBGs deviate from
the relations seen in Fig.\ \ref{sigcorr.plot}.

It is still unclear how \orgs\ fit in the general picture of galaxy
formation. They are redder, more metal rich, and more massive than
$z\approx 3$ Lyman break galaxies of the same luminosity, which
suggests they will evolve into more massive galaxies. A plausible
scenario is that they are descendants of LBGs at $z\sim 5$ or beyond,
in which continued star formation led to a gradual build-up of dust,
metals, and stellar mass.  It is tempting to place \orgs\ in an
evolutionary sequence linking them to Extremely Red Objects (EROs)
and massive $K-$selected galaxies at
$1<z<2$ (e.g., {Elston}, {Rieke}, \& {Rieke} 1988; {McCarthy} {et~al.} 2001; {Cimatti} {et~al.} 2002; Glazebrook et al.\ 2004) and
early-type galaxies in the local universe. The apparently smooth
sequence of massive galaxies in the $M/L$ versus redshift plane is
suggestive of such evolution. However, it is not clear whether
galaxies follow ``parallel tracks'' in such plots (i.e., \org\ $\rightarrow$
ERO $\rightarrow$ early-type, and LBG $\rightarrow$ spiral) or evolve
in complex ways due to
mergers or later addition of metal-poor gas
(e.g., Trager et al.\ 2000; van Dokkum \& Franx 2001; Bell et al.\ 2004).

The subsequent evolution of \orgs\ should also be placed in context
with the evolution of the cosmic stellar mass density.  The red galaxies
likely contribute $\gtrsim 50$\,\% of the stellar mass density at
$z\approx 2.5$ ({Franx} {et~al.} 2003; {Rudnick} {et~al.} 2003), a similar contribution
as that of elliptical galaxies to the local stellas mass density
(e.g., {Fukugita}, {Hogan}, \&  {Peebles} 1998). However, studies of
the Hubble Deep Fields suggest that the stellar mass density of the
Universe has increased by a factor of $\sim 10$ since $z\approx 3$
({Dickinson} {et~al.} 2003; {Rudnick} {et~al.} 2003), and it is unknown whether
\orgs\ participated in this rapid evolution.

The high dust content that we infer for \orgs\  has important
consequences for tests of galaxy formation models that use
the abundance of $K$-selected galaxies at high redshift
(e.g., Kauffmann \& Charlot 1998; Cimatti et al.\ 2003).
Our results show that any observed decline in the abundance will
at least in part be due to dust: the most luminous $K-$selected
galaxies at $0<z<1$ are dust-free elliptical galaxies, but the
$K$ magnitudes of red galaxies
at $z>2$ are probably attenuated by 1--2 magnitudes.
Therefore, tests of
galaxy formation models using the ``K20'' technique require a careful
treatment of dust, either by correcting the observed abundance for
extinction or by incorporating dust in the models.

There are several ways to establish the properties of red $z>2$
galaxies more firmly, and to enlarge the sample for improved statistics.
Photometric surveys over larger areas and multiple lines of sight
are needed to better quantify the surface density, clustering,
and to measure the luminosity function. Furthermore,
the high dust content and star formation rates imply that a
significant fraction of \orgs\ may shine brightly in the submm, as
evidenced by the SCUBA detection of the galaxy with the highest dust
content in our sample. 
The star formation may also be detectable with
Chandra in stacked exposures, as has been demonstrated recently for
LBGs (Reddy \& Steidel 2004).  Chandra can also provide estimates of
the prevalence of AGN in \orgs; this fraction could be relatively high
if black hole accretion rates correlate with stellar mass at early
epochs. Spitzer can provide better constraints on the
stellar masses, as it samples the rest-frame $K$ band.

Finally,
the results in this paper can clearly be improved by obtaining more
NIR spectroscopy of red $z>2$ galaxies. Contrary to the situation
for LBGs, the sample
of \orgs\ with measured redshift is rather limited due
to their faintness in the rest-frame UV and the resulting
difficulty of measuring redshifts in the observer's optical
(see van Dokkum et al.\ 2003; Wuyts et al.\ in preperation).
Furthermore, samples
of galaxies
with optically-measured redshifts are obviously biased toward galaxies
with high star formation rates, low dust content, and/or active nuclei;
it may not be a coincidence that the galaxy with the highest dust content
is the only one in our sample without Ly$\alpha$ emission.
``Blind'' NIR spectroscopy is notoriously difficult, but may be the
best way of determining the masses, star formation rates, and dust
content of \orgs\ in an unbiased fashion. It is expected that such
studies will be possible for large samples when multi-object
NIR spectrographs become available on 8--10m class telescopes.

\acknowledgements{
We thank Alice Shapley
for providing transmission curves of the $U_nG{\cal R}$ filters,
and Chuck Steidel, Dawn Erb, and Richard Ellis for their comments
and suggestions. We thank the anonymous referee for comments
that improved the paper.
The Caltech Time Allocation Committee is thanked for their
generous support for this program.
Support from the National Aeronautics and
Space Administration under grant NNG04GE12G, issued through the
Office of Space Science, is gratefully acknowledged.
The authors wish to recognize and acknowledge the
cultural role and reverance that the summit of Mauna Kea has
always had within the indigenous Hawaiian community. We are most
fortunate to have the opportunity to conduct observations from
this mountain.
}


\begin{appendix}
\section{Optical Spectroscopy in CDF-S}
\label{optspec.sec}

We obtained rest-frame UV spectroscopy of $J_s-K_s$ selected
galaxies in the Chandra Deep Field South (CDF-S) prior to our NIRSPEC
observations. The spectra were obtained with the Focal Reducer/Low
Dispersion Spectrograph 2 (FORS2) on the Very Large Telescope (VLT)
2002 December 5--6.  Galaxies with $J_s-K_s>2.3$ and $K_s<22$ were
selected using NIR photometry obtained from the public GOODS
survey (see \S \ref{cdfphot.sec}).
The 300V grism was used, giving an approximate
wavelength range of 0.3--1.0\,$\mu$.  Conditions were photometric, and
the seeing was $\approx 1\farcs 5$.  After each individual 1800\,s exposure the
objects were moved along the slit to facilitate sky subtraction.  A
bright blue star was included in each of the masks to enable
correction for atmospheric absorption.
The reduction followed standard procedures for dithered multi-slit
spectroscopic data (see, e.g., van Dokkum \& Stanford 2003).
An initial inspection of the reduced data in 2002 December
provided two redshifts for $J_s-K_s$ selected objects; as these
galaxies were subsequently
observed with NIRSPEC their spectra are
presented here. The full photometric and spectroscopic
datasets of $J_s-K_s$ selected galaxies in CDF-S
will be presented in Wuyts et al., in preparation.

The two spectra are shown in Fig.\ \ref{optspec.plot}.
Galaxy C--237 has a strong emission line at
4354\,\AA. The line is identified as Ly$\alpha$ redshifted
to $z=2.5808$,
as no other emission lines are seen in the optical window.
The spectrum of galaxy C--3396
is qualitatively similar to spectra of LBGs
(see, e.g., Steidel et al.\ 1996),
and the galaxy has an unambiguous redshift $z=3.25$.
The two strongest
absorption lines are blueshifted with respect to Ly$\alpha$,
as is common in LBGs (Shapley et al.\ 2003).
We find $z_{\rm Ly\alpha}=3.2538$,
$z_{\rm Si\,{II}}=3.2462$, and $z_{\rm C\,{II}}=3.2458$,
each with an error of $\approx 0.0005$.

\vbox{
\begin{center}
\leavevmode
\hbox{%
\epsfxsize=8cm
\epsffile{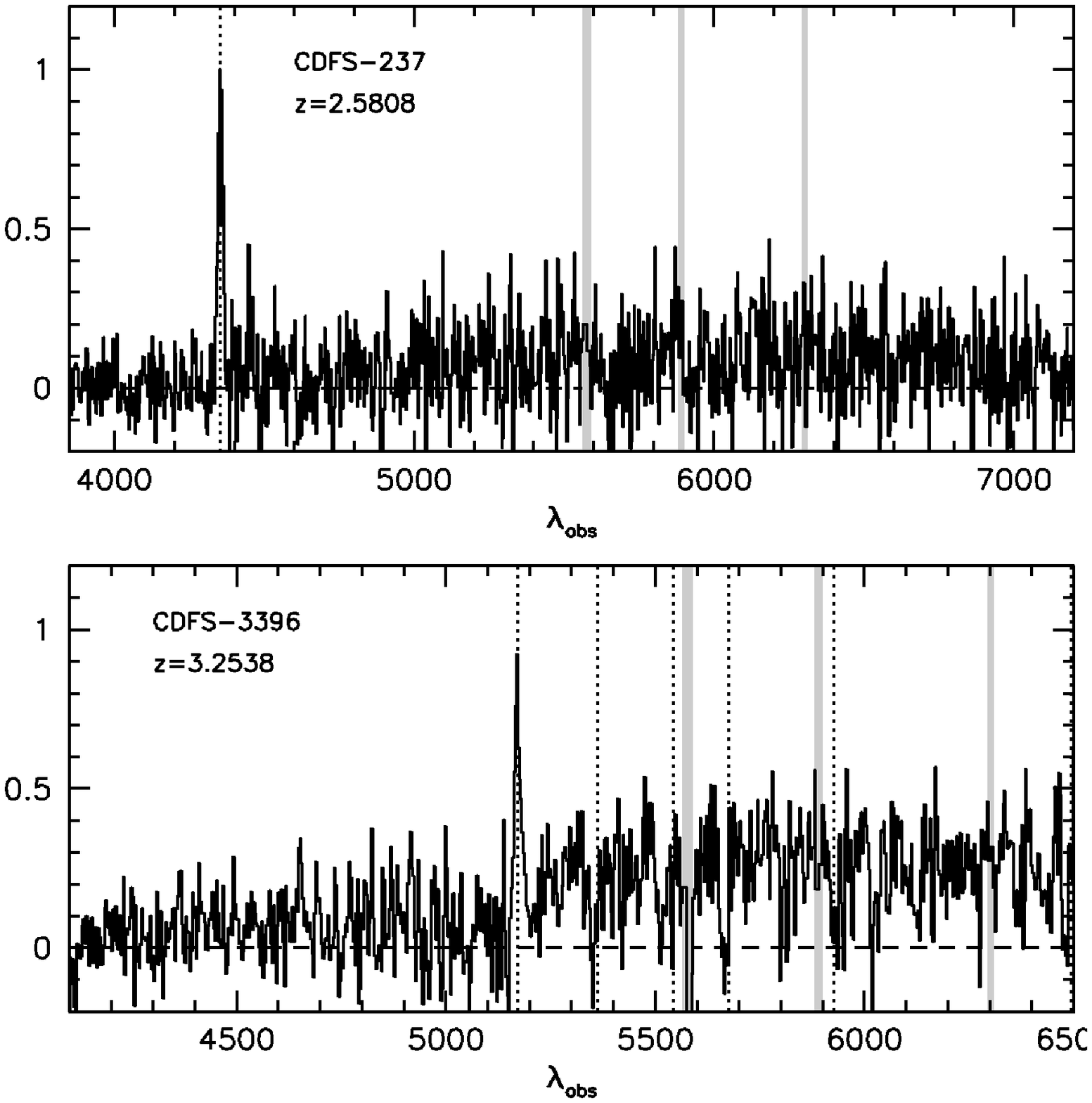}}
\figcaption{\small
Rest frame UV spectra of two galaxies with $J_s-K_s>2.3$ in CDF-S.
Grey bands indicate locations of strong atmospheric emission lines.
Galaxy C--237 has a weak continuum and one strong emission line
which we interpret as Ly-$\alpha$ at $z=2.5808$. Galaxy
C--3396 has Ly-$\alpha$ in emission and shows interstellar
absorption lines typical of star forming galaxies.
\label{optspec.plot}}
\end{center}}

\section{Are LBGs and DRGs Orthogonal Populations?}

The fact that the majority of $J_s-K_s>2.3$ galaxies are fainter than
$R=25.5$ in the rest-frame ultraviolet provides an upper limit to
the number of galaxies that may fall in both categories. {Franx} {et~al.} (2003)
and {van Dokkum} {et~al.} (2003) find that more than 70\,\% or the red galaxies
are fainter than this limit, and would therefore not be present in
spectroscopic samples. Here we examine whether the most luminous
\orgs\ (which are the subject of this study) would be selected by
the Lyman break technique.

The redshift information in addition to the high quality SEDs allow us
to determine whether the bright galaxies in our sample would
be selected as Lyman-break galaxies. Synthetic colors in the
Steidel \& Hamilton (1992)
filter system are determined from the best fitting Bruzual \& Charlot
(2003) models, including extinction and absorption due to the
Lyman limit and the Ly$\alpha$ forest (Madau et al.\ 1996).
In Fig.\ \ref{lbgcomp.plot} we
show the locations of the seven galaxies in the $U_n-G$ versus
$G-{\cal R}$ plane. None of the seven \orgs\ with spectroscopic
redshift fall in the
``C+D'' selection region; two are just inside
the wider ``M+MD'' region (see {Steidel} {et~al.} 2003).

\vbox{
\begin{center}
\leavevmode
\hbox{%
\epsfxsize=7.5cm
\epsffile{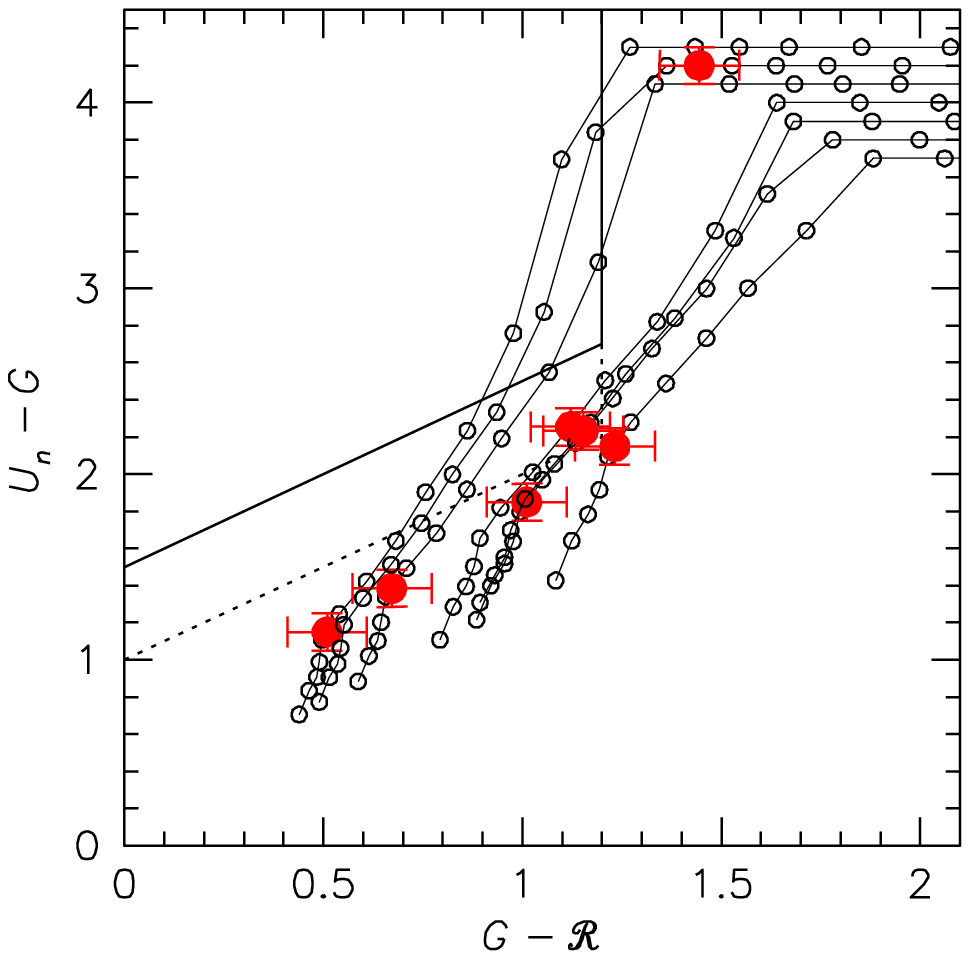}}
\figcaption{\small
Location of the $J_s-K_s$ selected galaxies on the Lyman-break
selection diagram. Solid symbols with errorbars show the rest-UV
colors of \orgs\ at their observed redshift. Tracks and open symbols show the
effect of varying the redshift over the interval $2<z<4$.
Solid lines delineate the primary selection
region (``C+D'') employed by Steidel and collaborators; the broken line
indicates their more generous ``M+MD'' criteria (see Steidel et al.\
2003). The $J_s-K_s$ objects
have red rest-frame UV spectra due to dust, and would only
be selected as LBGs in a very small redshift range, or not at all.
\label{lbgcomp.plot}}
\end{center}}

This result partly
reflects the narrow redshift range that the
Steidel \& Hamilton filters are tuned to:
Lyman-break galaxies at $z=2.4$ would require a ``bluer'' filter set
than $U_nG{\cal R}$, or a different selection window in the
$U_n-G$ vs.\ $G-{\cal R}$ diagram (see Erb et al.\ 2003).
Rather than blueshifting the filters we artificially redshift
the galaxy spectra over the range $2<z<4$, properly accounting for
the changing intergalactic absorption. As can be seen in
Fig.\ \ref{lbgcomp.plot} three of the seven galaxies would fall in
the primary selection area if they were at $z\approx 3$.
The $G-{\cal R}$ colors of the remaining four are too red
to fall in the primary selection area for any redshift; however,
three are just inside the wider ``M+MD'' area for $z\approx 2.7$.

Even though six out of seven galaxies could be selected as ``C+D'' or
``M+MD'' Lyman breaks for certain redshifts, the probability that they
enter spectroscopic LBG samples is low. For a random distribution in
redshift, star forming galaxies with $E(B-V)=0.0-0.3$ are $\sim 5$
times more likely to fall in the selection box than the dusty
\orgs\ (see Fig.\ 2 in Steidel et al.\ 2003).
Based on our sample of seven we conclude that even the small fraction
of red galaxies which has $R<25.5$ is underrepresented in
spectroscopic LBG samples, because of their red rest-frame UV colors.

\end{appendix}

\newpage

\begin{small}
\begin{center}
{ {\sc TABLE 1} \\
\sc Galaxy Sample}\\
\vspace{0.1cm}
\begin{tabular}{llcccccc}
\hline
\hline
Id & vD03$^a$ & $z_{\rm UV}$ & $K_s^b$ & $K_{s, \rm corr}^{c}$ & $J_s-K_s$ &  Exp\,(s)& $\lambda\,(\mu$m) \\
\hline
M--1035 & 1195 & 2.424     & 19.57      &  19.91   &  2.25    & 3,600 & 2.058 -- 2.475 \\
     &      &           &       &     &        & 1,800 & 1.508 -- 1.778 \\
     &      &           &       &     &       & 2,700 & 1.266 -- 1.540 \\
M--1356 & 1458 & 2.427     & 19.67       &  19.82   &  2.13       & 3,600 & 2.058 -- 2.480 \\
M--1383 & 1656 & 2.423$^d$     &  19.55     &    19.77 &   2.83      & 5,400 & 1.993 -- 2.410 \\
M--1319 & 1671 & 2.423     &    19.01   &  19.15   &  2.58      &  5,400 & 2.050 -- 2.470 \\
     &      &           &       &     &        &  3,600 & 1.282 -- 1.558 \\
C--237 & --- & 2.581 & 20.00 & --- & 2.63 &    5,400 & 2.056 -- 2.480 \\
M--140 & 184  & 2.705 &  19.76 & 19.87 & 2.68 & --- & --- \\
C--3396 &--- & 3.254 & 19.66 & --- & 2.56 &    4,500 & 1.980 -- 2.400 \\
\hline
\end{tabular}
\end{center}
{\small
$^a$\,Id number from van Dokkum et al.\ (2003).\\
$^b$\,Total $K_s$ magnitude on Vega system.\\
$^c$\,$K_s$ magnitude corrected for the lensing effect of the foreground
cluster.\\
$^d$\,Revised redshift from rest-frame optical spectroscopy (see
\S\,\ref{detect.sec}).\\
}
\end{small}

\begin{small}
\begin{center}
{ {\sc TABLE 2} \\
\sc Emission Lines}\\
\vspace{0.1cm}
\begin{tabular}{lclcccc}
\hline
\hline
Id & $z_{\rm opt}$ & Line & $F^a$ & $W^b$ & $L^c$ & $\sigma^d$\\
\hline
M--1035 & $2.424$    & H$\beta$ & $2.9 \pm 2.1$ & $28 \pm 20$ & $1.4 \pm 1.0$& --- \\
        &            & \o4959\ & $13 \pm 4$ & $116 \pm 42$ & $5.9 \pm 1.8$ &
  $231 \pm 26$ \\
        &            & \5007\ & $31 \pm 7$ & $275 \pm 86$ & $14.3 \pm 3.2$ &
  $231 \pm 26$\\
        &            & H$\alpha$\,+\,[N\,{\sc ii}] & $17 \pm 4$ & $114 \pm 32$ & $8.1 \pm 1.9$ &
 $\lesssim 1600$\\
M--1356 & $2.430$    & H$\alpha$\,+\,[N\,{\sc ii}] & $19 \pm 4$ & $104 \pm 23$ & $8.7 \pm 2.0$ & $\lesssim 2500$\\
M--1383 & $2.423$    & H$\alpha$ & $6.5 \pm 1.7$ & $31 \pm 9$ & $3.0 \pm 0.8$ &
 $153 \pm 25$ \\
        &            & [N\,{\sc ii}] & $2.6 \pm 0.9$ & $13\pm 4$ & $1.2 \pm 0.4$
 & $153 \pm 25$ \\
M--1319 & $2.424$    & H$\alpha$ & $6.3 \pm 1.4$ & $19 \pm 4$ & $2.9 \pm 0.7$ &
 $240 \pm 56$\\
        &            & [N\,{\sc ii}] & $5.0 \pm 1.2$ & $15 \pm 3$ &$2.3\pm 0.5$&
 $240 \pm 56$\\
C--237  & $2.581$ & H$\alpha$\,+\,[N\,{\sc ii}]
& $7.9 \pm 2.4$ & $46 \pm 11$ & $4.3 \pm 1.3$ & $\lesssim 1000$\\
C--3396 & $3.249$ & H$\beta$ & $2.5^{+2.5}_{-1.2}$ & $7^{+7}_{-4}$ & $2.4^{+2.4}_{-1.2}$ & ---\\
           &         & \o4959\ & $4.5 \pm 2.1$ & $12\pm 5$ & $4.3 \pm 2.0$ &
 $134_{-46}^{+87}$ \\
           &         & \5007\ & $10 \pm 3$ & $26\pm 8$ & $9.5 \pm 2.9$ &
 $134_{-46}^{+87}$\\
\hline
\end{tabular}
\end{center}
{\small $^a$\,Line flux in units of $10^{-17}$\,ergs\,s$^{-1}$\,cm$^{-2}$.\\
$^b$\,Rest-frame equivalent width in \AA.\\
$^c$\,Line luminosity in units of $10^{42}$\,ergs\,s$^{-1}$.\\
$^d$\,Line width in \kms.\\
}
\end{small}

\begin{small}
\begin{center}
{ {\sc TABLE 3} \\
\sc Emisson Line Corrections}$^a$\\
\vspace{0.1cm}
\begin{tabular}{lccccccccc}
\hline
\hline
Id & $U$ & $B$ & $V$ & $V_{606}$ & $I_{814}$ &
$z$ & $J_s$ & $H$ & $K_s$ \\
\hline
M--1035   & 0  & $1.2\pm 0.3$ & $0.56\pm 0.20$ & $0.39\pm 0.10$
& $<0.10$ & ---
& $0.11 \pm 0.10$ & $0.31 \pm 0.06$ & $0.16 \pm 0.03$ \\
M--1356   &  0 & $0.12 \pm 0.02$ & $0.09 \pm 0.02$ & $0.02 \pm 0.02$
& $<0.05$ & --- & $(0.07 \pm 0.03)$ & $(0.24 \pm 0.10)$ & $0.16 \pm 0.02$ \\
M--1383   & 0 & $<0.05$ & $<0.05$ & $<0.05$ & 
$<0.05$ & --- & $(0.06 \pm 0.03)$
& $(0.03 \pm 0.02)$ & $0.08 \pm 0.02$ \\
M--1319   & 0 & $0.10 \pm 0.02$ & $<0.05$ & $<0.05$ & $<0.05$ & --- &
$(0.03 \pm 0.03)$ & $(0.02 \pm 0.02)$ & $0.04 \pm 0.02$ \\
C--237  & 0 & $0.77 \pm 0.20$ & $<0.05$ & $<0.05^b$ & $<0.05^c$ &
 $(<0.05)$ & $(<0.1)$ & $(<0.1)$ & 0 \\
M--140    & 0 & $0.26 \pm 0.03$ & $<0.05$ & $<0.05$  & $<0.05$ & ---
& $(0.05 \pm 0.05)$ & $(<0.05)$ & $(<0.05)$ \\
C--3396 & 0 & 0 & $0.06 \pm 0.02$ & $<0.05^b$ & $<0.05^c$ & $(<0.05)$ &
$(<0.05)$ & $(0.08 \pm 0.07)$ & $0.07 \pm 0.05$ \\
\hline
\end{tabular}
\end{center}
{\small $^a$\,Units are magnitudes; values in parentheses were not
measured but calculated based on line ratios of nearby
galaxies.\\
$^b$\,FORS $R$ filter.\\
$^c$\,FORS $I$ filter.\\
}
\end{small}


\newpage
\begin{small}
\begin{center}
{ {\sc TABLE 4} \\
\sc Results From SED Fits}\\
\vspace{0.1cm}
\begin{tabular}{lcccc}
\hline
\hline
Id & $\log \tau$ & $A_V$ & $\log$\,SFR & $\log M_*$  \\
  & yr & mag & $M_{\odot}\,{\rm yr}^{-1}$ & $M_{\odot}$ \\
\hline
M--1035  & $8.9^{+0.2}_{-0.2}$ & $1.9^{+0.1}_{-0.2}$ & $2.50_{-0.10}^{+0.04}$
& $11.26_{-0.13}^{+0.17}$ \\
M--1356  & $9.2^{+0.2}_{-0.2}$ & $0.9^{+0.1}_{-0.1}$ & $2.01_{-0.10}^{+0.10}$
& $11.13_{-0.12}^{+0.09}$ \\
M--1383  & $9.0^{+0.4}_{-0.6}$ & $2.3^{+0.5}_{-0.3}$ & $2.64_{-0.20}^{+0.36}$
& $11.55_{-0.21}^{+0.14}$ \\
M--1319  & $9.4^{+0}_{-0.3}$ & $1.3^{+0.1}_{-0.1}$ & $2.39_{-0.04}^{+0.09}$
& $11.68_{-0.08}^{+0.08}$ \\
C--237  & $9.4^{+0}_{-0.4}$ & $2.0^{+0.2}_{-0.1}$ & $2.40_{-0.08}^{+0.16}$
& $11.69_{-0.17}^{+0.08}$ \\
M--140  & $8.9^{+0.2}_{-0.2}$ & $1.8^{+0.1}_{-0.1}$ & $2.67_{-0.09}^{+0.08}$
& $11.43_{-0.09}^{+0.14}$ \\
C--3396  & $9.2^{+0.1}_{-0.6}$ & $1.1^{+0.2}_{-0.2}$ & $2.55_{-0.13}^{+0.21}$ & $11.60_{-0.34}^{+0.07}$ \\
\hline
\end{tabular}
\end{center}
\end{small}

\begin{small}
\begin{center}
{ {\sc TABLE 5} \\
\sc Star Formation Rates in Different Models}\\
\vspace{0.1cm}
\begin{tabular}{l|cc|cc}
\hline
\hline
 & \multicolumn{2}{|c|}{Continuous ($\tau=\infty$)} &
\multicolumn{2}{c}{Declining ($\tau=300$\,Myr)} \\
 & M--1383 & M--1319 & M--1383 & M--1319 \\
\hline
Age\,(SED) (Gyr)$^a$ & 1.0 & 2.6 & 0.7 & 0.7 \\
$A_V$\,(SED)$^a$ & 2.3 & 1.3 & 1.8 & 1.0 \\
SFR\,(SED)$^a$ & 439 & 243 & 122 & 106 \\
SFR\,(H$\alpha$; $A_{\rm H\,{II}}=0$)$^b$ & 24 & 23 & 24 & 23 \\
SFR\,(H$\alpha$; $A_{\rm H\,{II}}=A_V$)$^b$ & 169 & 69 & 110 & 54 \\
SFR\,(H$\alpha$; $A_{\rm H\,{II}}=A_V + 1$)$^b$ & 393 & 161 &  257 & 125 \\
\hline
\end{tabular}
\end{center}
{\small $^a$\,Ages, star formation rates and extinction derived from fits to
broad band spectral energy distributions.\\
$^b$\,Star formation rates derived from H$\alpha$, for different
assumptions for the extinction toward H\,{\sc ii} regions.\\
}
\end{small}


\begin{references}

\reference{} {Adelberger}, K.~L. \& {Steidel}, C.~C. 2000, \apj, 544, 218

\reference{} {Adelberger}, K.~L., {Steidel}, C.~C., {Shapley}, A.~E., \& {Pettini}, M. 2003,  \apj, 584, 45

\reference{} Alexander, D.~M., et al.\ 2002, \aj, 122, 2156

\reference{} {Armus}, L., {Heckman}, T.~M., \& {Miley}, G.~K. 1989, \apj, 347, 727

\reference{} {Barger}, A.~J., {Cowie}, L.~L., {Mushotzky}, R.~F., \& {Richards}, E.~A. 2001,  \aj, 121, 662

\reference{} {Baugh}, C.~M., {Cole}, S., {Frenk}, C.~S., \& {Lacey}, C.~G. 1998, \apj, 498,  504

\reference{} {Beers}, T.~C., {Flynn}, K., \& {Gebhardt}, K. 1990, \aj, 100, 32

\reference{} {Blain}, A.~W., {Smail}, I., {Ivison}, R.~J., {Kneib}, J.-P., \& {Frayer},  D.~T. 2002, \physrep, 369, 111

\reference{} {Bolzonella}, M., {Miralles}, J.-M., \& {Pell{\' o}}, R. 2000, \aap, 363, 476

\reference{} {Bruzual}, G. \& {Charlot}, S. 2003, \mnras, 344, 1000

\reference{} {Calzetti}, D., {Armus}, L., {Bohlin}, R.~C., {Kinney}, A.~L., {Koornneef}, J.,  \& {Storchi-Bergmann}, T. 2000, \apj, 533, 682

\reference{} {Cimatti}, A., {Daddi}, E., {Mignoli}, M., {Pozzetti}, L., {Renzini}, A.,  {Zamorani}, G., {Broadhurst}, T., {Fontana}, A., {et al.} 2002, \aap, 381, L68

\reference{} {Daddi}, E., {Cimatti}, A., {Renzini}, A., {Vernet}, J., {Conselice}, C.,  {Pozzetti}, L., {Mignoli}, M., {Tozzi}, P., {et al.} 2004, \apjl, 600, L127

\reference{} {Daddi}, E., {R{\" o}ttgering}, H.~J.~A., {Labb{\' e}}, I., {Rudnick}, G.,  {Franx}, M., {Moorwood}, A.~F.~M., {Rix}, H.~W., {van der Werf}, P.~P., {et al.} 2003, \apj, 588, 50

\reference{} {Denicol{\' o}}, G., {Terlevich}, R., \& {Terlevich}, E. 2002, \mnras, 330, 69

\reference{} {Dickinson}, M., {Papovich}, C., {Ferguson}, H.~C., \& {Budav{\' a}ri}, T.  2003, \apj, 587, 25

\reference{} {Eggen}, O.~J., {Lynden-Bell}, D., \& {Sandage}, A.~R. 1962, \apj, 136, 748

\reference{} {Elston}, R., {Rieke}, G.~H., \& {Rieke}, M.~J. 1988, \apjl, 331, L77

\reference{} {Erb}, D.~K., {Shapley}, A.~E., {Steidel}, C.~C., {Pettini}, M., {Adelberger},  K.~L., {Hunt}, M.~P., {Moorwood}, A.~F.~M., \& {Cuby}, J. 2003, \apj, 591,  101

\reference{} F\"orster Schreiber, N.~M., et al. 2004, \apj, submitted

\reference{} Francis, P.~J., Woodgate, B.~E., \& Danks, A.~C. 1997,
ApJ, 482, L25

\reference{} {Franx}, M., {Illingworth}, G.~D., {Kelson}, D.~D., {van Dokkum}, P.~G., \&  {Tran}, K. 1997, \apjl, 486, L75

\reference{} {Franx}, M., {Labb{\' e}}, I., {Rudnick}, G., {van Dokkum}, P.~G., {Daddi}, E.,  {F{\" o}rster Schreiber}, N.~M., {Moorwood}, A., {Rix}, H., {et al.} 2003, \apjl, 587, L79


\reference{} {Fukugita}, M., {Hogan}, C.~J., \& {Peebles}, P.~J.~E. 1998, \apj, 503, 518

\reference{} Giacconi, R., et al.\ 2002, \apjs, 139, 369

\reference{} {Giavalisco}, M., {Steidel}, C.~C., {Adelberger}, K.~L., {Dickinson}, M.~E.,  {Pettini}, M., \& {Kellogg}, M. 1998, \apj, 503, 543

\reference{} Glazebrook, K., et al. 2004, ApJ, submitted (astro-ph/0401037)

\reference{} {Green}, P.~J. \& {Mathur}, S. 1996, \apj, 462, 637

\reference{} Hall, P., et al.\ 2001, AJ, 121, 1840

\reference{} {Hoekstra}, H., {Franx}, M., \& {Kuijken}, K. 2000, \apj, 532, 88

\reference{} {Hu}, E.~M., {Cowie}, L.~L., \& {McMahon}, R.~G. 1998, \apjl, 502, L99+

\reference{} {Im}, M., {Yamada}, T., {Tanaka}, I., \& {Kajisawa}, M. 2002, \apjl, 578, L19

\reference{} {Jansen}, R.~A., {Fabricant}, D., {Franx}, M., \& {Caldwell}, N. 2000, \apjs,  126, 331

\reference{} {Johnson}, O., {Best}, P.~N., \& {Almaini}, O. 2003, \mnras, 343, 924

\reference{} {Kauffmann}, G., {Colberg}, J.~M., {Diaferio}, A., \& {White}, S.~D.~M. 1999,  \mnras, 303, 188

\reference{} {Kauffmann}, G., {White}, S.~D.~M., \& {Guiderdoni}, B. 1993, \mnras, 264, 201

\reference{} {Kennicutt}, R.~C. 1998, \araa, 36, 189

\reference{} {Kewley}, L.~J., {Geller}, M.~J., {Jansen}, R.~A., \& {Dopita}, M.~A. 2002,  \aj, 124, 3135

\reference{} {Kobulnicky}, H.~A., {Kennicutt}, R.~C., \& {Pizagno}, J.~L. 1999, \apj, 514,  544

\reference{} {Labb{\' e}}, I., {Franx}, M., {Rudnick}, G., {Schreiber}, N.~M.~F., {Rix}, H.,  {Moorwood}, A., {van Dokkum}, P.~G., {van der Werf}, P., {et al.} 2003, \aj, 125, 1107

\reference{} {Madau}, P., {Ferguson}, H.~C., {Dickinson}, M.~E., {Giavalisco}, M.,  {Steidel}, C.~C., \& {Fruchter}, A. 1996, \mnras, 283, 1388

\reference{} {McCarthy}, P.~J., {Carlberg}, R.~G., {Chen}, H.-W., {Marzke}, R.~O., {Firth},  A.~E., {Ellis}, R.~S., {Persson}, S.~E., {McMahon}, R.~G., {et al.} 2001, \apjl, 560, L131

\reference{} {Meza}, A., {Navarro}, J.~F., {Steinmetz}, M., \& {Eke}, V.~R. 2003, \apj, 590,  619

\reference{} {Norman}, C., {Hasinger}, G., {Giacconi}, R., {Gilli}, R., {Kewley}, L.,  {Nonino}, M., {Rosati}, P., {Szokoly}, G., {et al.} 2002, \apj, 571, 218

\reference{} {Pagel}, B.~E.~J., {Edmunds}, M.~G., {Blackwell}, D.~E., {Chun}, M.~S., \&  {Smith}, G. 1979, \mnras, 189, 95

\reference{} {Papovich}, C., {Dickinson}, M., \& {Ferguson}, H.~C. 2001, \apj, 559, 620

\reference{} {Pettini}, M., {Kellogg}, M., {Steidel}, C.~C., {Dickinson}, M., {Adelberger},  K.~L., \& {Giavalisco}, M. 1998, \apj, 508, 539

\reference{} {Pettini}, M., {Shapley}, A.~E., {Steidel}, C.~C., {Cuby}, J., {Dickinson}, M.,  {Moorwood}, A.~F.~M., {Adelberger}, K.~L., \& {Giavalisco}, M. 2001, \apj,  554, 981

\reference{} {Pettini}, M., {Steidel}, C.~C., {Adelberger}, K.~L., {Dickinson}, M., \&  {Giavalisco}, M. 2000, \apj, 528, 96

\reference{} {R{\" o}ttgering}, H., {Daddi}, E., {Overzier}, R., \& {Wilman}, R. 2003, New  Astronomy Review, 47, 309

\reference{} {Riess}, A.~G., {}, {}, \& {ETAL}. 1998, \aj, 116, 1009

\reference{} {Rix}, H., {Guhathakurta}, P., {Colless}, M., \& {Ing}, K. 1997, \mnras, 285,  779

\reference{} {Rudnick}, G., {Rix}, H., {Franx}, M., {Labb{\' e}}, I., {Blanton}, M.,  {Daddi}, E., {F{\" o}rster Schreiber}, N.~M., {Moorwood}, A., {et al.} 2003, \apj, 599, 847

\reference{} {Shapley}, A.~E., {Steidel}, C.~C., {Adelberger}, K.~L., {Dickinson}, M.,  {Giavalisco}, M., \& {Pettini}, M. 2001, \apj, 562, 95

\reference{} {Spergel}, D.~N., {Verde}, L., {Peiris}, H.~V., {Komatsu}, E., {Nolta}, M.~R.,  {Bennett}, C.~L., {Halpern}, M., {Hinshaw}, G., {et al.} 2003, \apjs, 148, 175

\reference{} Stiavelli, M., Scarlata, C., Panagia, N., Treu, T., Bertin, G.,
\& Bertola, F. 2001, ApJ, 561, L37

\reference{} {Steidel}, C.~C., {Adelberger}, K.~L., {Giavalisco}, M., {Dickinson}, M., \&  {Pettini}, M. 1999, \apj, 519, 1

\reference{} {Steidel}, C.~C., {Adelberger}, K.~L., {Shapley}, A.~E., {Pettini}, M.,  {Dickinson}, M., \& {Giavalisco}, M. 2003, \apj, 592, 728

\reference{} {Steidel}, C.~C., {Giavalisco}, M., {Pettini}, M., {Dickinson}, M., \&  {Adelberger}, K.~L. 1996, \apjl, 462, L17

\reference{} {Steidel}, C.~C. \& {Hamilton}, D. 1992, \aj, 104, 941

\reference{} {Steidel}, C.~C., {Pettini}, M., \& {Hamilton}, D. 1995, \aj, 110, 2519

\reference{} {Stern}, D., {Moran}, E.~C., {Coil}, A.~L., {Connolly}, A., {Davis}, M.,  {Dawson}, S., {Dey}, A., {Eisenhardt}, P., {et al.} 2002, \apj,  568, 71

\reference{} {Storchi-Bergmann}, T., {Calzetti}, D., \& {Kinney}, A.~L. 1994, \apj, 429, 572

\reference{} Trujillo, I., et al.\ 2004, ApJ, 604, 521

\reference{} {Tully}, R.~B. \& {Fisher}, J.~R. 1977, \aap, 54, 661

\reference{} {van der Marel}, R.~P. 1991, \mnras, 253, 710

\reference{} {van Dokkum}, P.~G. \& {Ellis}, R.~S. 2003, \apjl, 592, L53

\reference{} {van Dokkum}, P.~G., {F{\" o}rster Schreiber}, N.~M., {Franx}, M., {Daddi}, E.,  {Illingworth}, G.~D., {Labb{\' e}}, I., {Moorwood}, A., {Rix}, H., {et al.} 2003, \apjl, 587, L83

\reference{} {van Dokkum}, P.~G. \& {Franx}, M. 1996, \mnras, 281, 985

\reference{} ---. 2001, \apj, 553, 90

\reference{} {van Dokkum}, P.~G., {Franx}, M., {Fabricant}, D., {Illingworth}, G.~D., \&  {Kelson}, D.~D. 2000, \apj, 541, 95

\reference{} {van Dokkum}, P.~G., {Franx}, M., {Kelson}, D.~D., \& {Illingworth}, G.~D.  2001, \apjl, 553, L39

\reference{} {van Dokkum}, P.~G. \& {Stanford}, S.~A. 2003, \apj, 585, 78

\reference{} {White}, S.~D.~M. \& {Frenk}, C.~S. 1991, \apj, 379, 52

\end{references}
\end{document}